\documentclass[numberedappendix]{emulateapj}

\usepackage{mystyle}

\shortauthors{Li et al.}
\shorttitle{CO Intensity Mapping in the Epoch of Galaxy Assembly}

\begin{document}

\title{Connecting CO Intensity Mapping to Molecular Gas and Star
  Formation \\
in the Epoch of Galaxy Assembly}

\author{Tony Y. Li$^1$, Risa H. Wechsler$^{1,2}$, Kiruthika Devaraj$^1$, Sarah E. Church$^1$}
\affil{$^1$Kavli Institute for Particle Astrophysics and Cosmology;\\
    Physics Department, Stanford University, Stanford, CA 94305,
    USA; \\
    {\tt tonyyli@stanford.edu, rwechsler@stanford.edu}}
\affil{$^2$    SLAC National Accelerator Laboratory, Menlo Park, CA 94025, USA}

\date{\today}

\begin{abstract}
\noindent
  Intensity mapping, which images a single spectral line from
  unresolved galaxies across cosmological volumes, is a promising
  technique for probing the early universe.  Here we present
  predictions for the intensity map and power spectrum of the CO(1-0)
  line from galaxies at $z \sim 2.4$--2.8, based on a parameterized model
  for the galaxy--halo connection, and demonstrate the extent to which
  properties of high-redshift galaxies can be directly inferred from
  such observations.  We find that our fiducial prediction should be
  detectable by a realistic experiment.  Motivated by significant
  modeling uncertainties, we demonstrate the effect on the power
  spectrum of varying each parameter in our model.  Using simulated
  observations, we infer constraints on our model parameter space with
  an MCMC procedure, and show corresponding constraints on the
  $\lir$--$\lco$ relation and the CO luminosity function.  These
  constraints would be complementary to current high-redshift galaxy
  observations, which can detect the brightest galaxies but not
  complete samples from the faint end of the luminosity function.  By
  probing these populations in aggregate, CO intensity mapping could
  be a valuable tool for probing molecular gas and its relation to
  star formation in high-redshift galaxies.
\end{abstract}

\maketitle

\section{Introduction}\label{sec:intro}

\begin{figure*}
    \centering
    \raisebox{-0.5\height}{
        \includegraphics[width=0.6\textwidth]{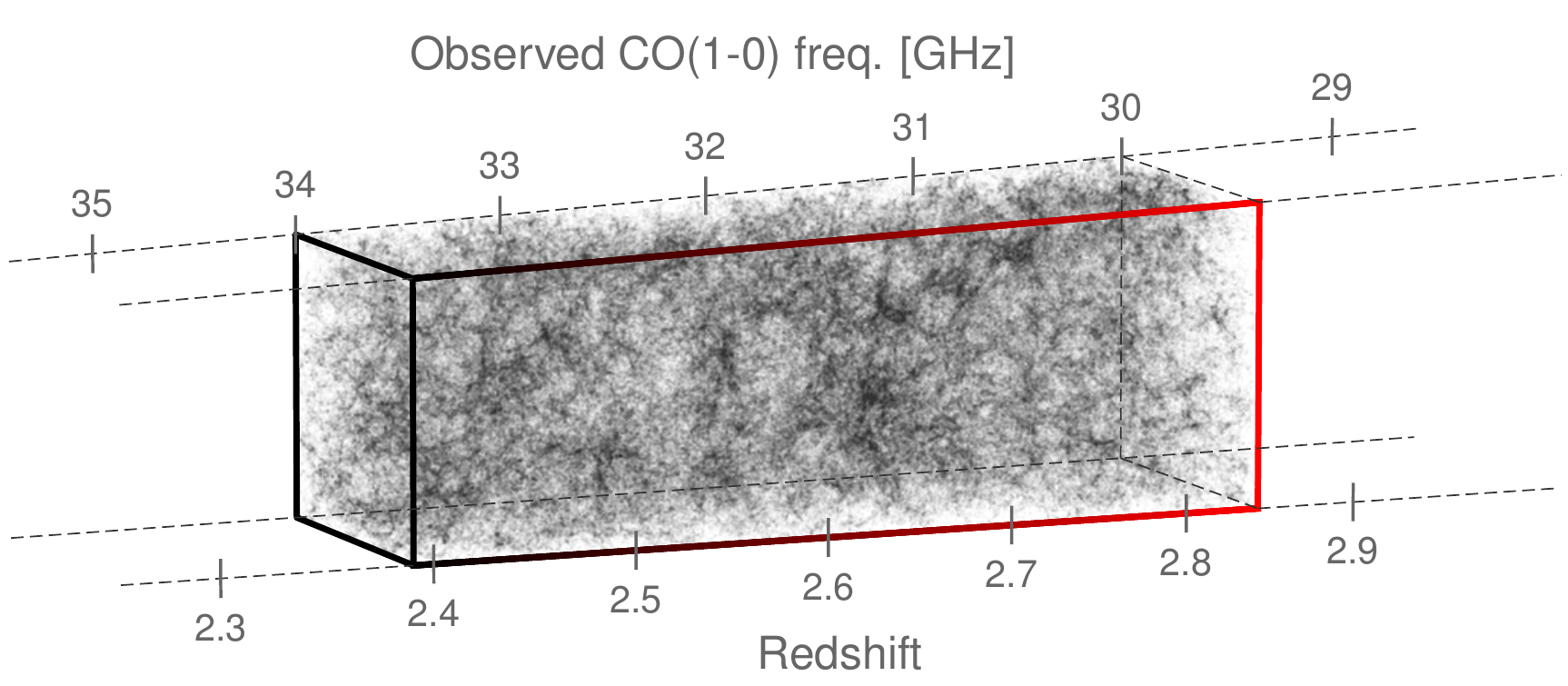}
    }
    \\
    \raisebox{-0.5\height}{
        \includegraphics[width=\textwidth]{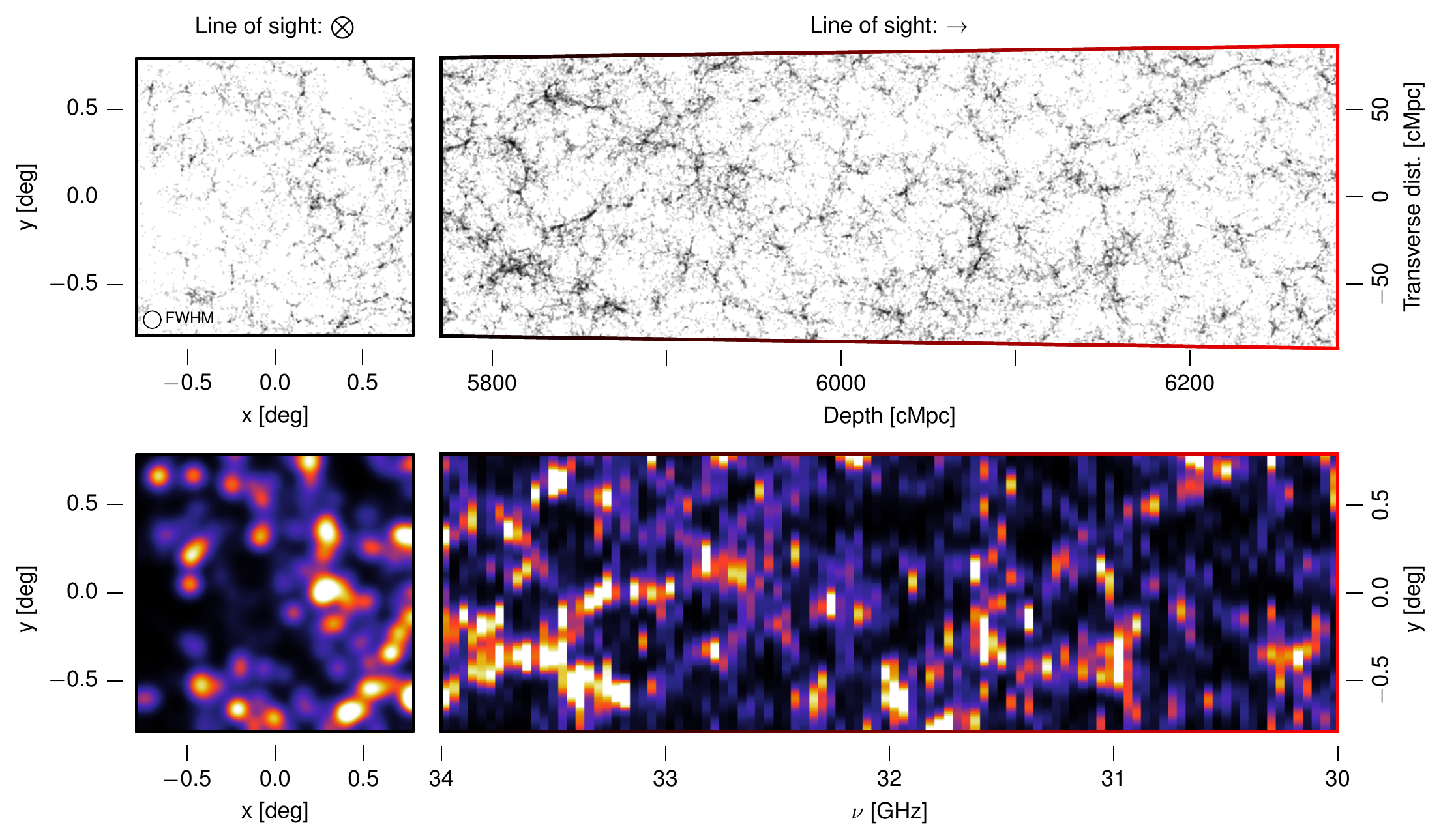}
    }
    \\
        \includegraphics[width=0.5\textwidth]{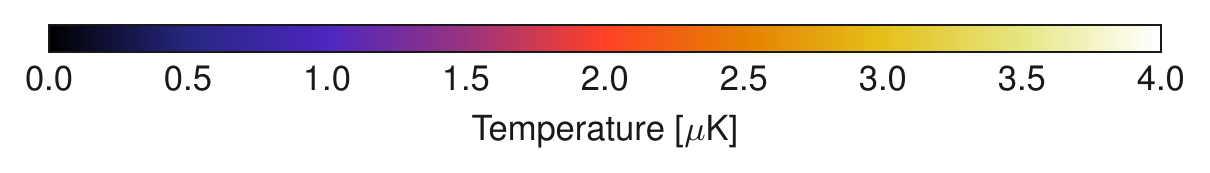}
    \caption{
        Input and output of our modeling process, i.e. initial dark matter halos and final CO intensity map (details in \S \ref{sec:comodel}).  These plots illustrate one realization of the pathfinder experiment's survey volume (\S \ref{sec:obsparams} and Table \ref{tab:obsparams}), while the full experiment's survey area is 2.5 times larger.
        \textbf{Top}: Halos in the 3D volume, rendered to scale in comoving distance.  Along the line-of-sight direction, we label the equivalent cosmological redshifts and redshifted CO(1-0) frequencies.
        \textbf{Middle}: 2D projections of halo positions.  The left image shows the ``front'' view of halos that would fall into the highest 40 MHz frequency channel, or lowest redshift slice.  The pathfinder beam size is shown for scale.  The right image shows the ``side'' view of halos to a depth of 6 arcmin, or one beam width.
        \textbf{Bottom}: CO intensity map produced by our fiducial model.  The slice volumes are the same as above, albeit with comoving depth converted to observed frequency.  The same large-scale structure is readily apparent in both images, even with the lower resolution of the intensity map.  The analysis in this paper relies on the power spectrum of this map (see Fig. \ref{fig:pspec_fid}).
    }
    \label{fig:halos2imap}
\end{figure*}

Current progress in cosmology and galaxy formation is strongly
informed by galaxy observations at increasingly high redshifts,
reaching into the epoch of reionization.  Modern observatories such as
ALMA may have even begun to probe the interstellar medium (ISM) of
``typical'' high-redshift galaxies
\citep[e.g.][]{riechers/etal:2014, watson/etal:2015}. Despite this, it remains difficult
and expensive to observe populations of such galaxies in statistically
large and complete samples, due to their faintness.

However, a full understanding of both cosmology and galaxy formation
depends on such samples.  To inform galaxy formation models,
individual galaxies need to be placed in the context of their broader
populations, for without such context it is unclear how representative
individually observed galaxies may be.  For cosmology, precise
clustering measurements that trace large-scale structure require
sufficient number densities of spectroscopically observed galaxies
\citep{blake/etal:2011, anderson/etal:2012}.

\emph{Intensity mapping} is a technique that can potentially address
this challenge in regimes that are inaccessible by typical galaxy
surveys.  The method aims to map a single spectral line across
cosmologically large volumes.  The angular resolution used to do this
images cumulative emission from multiple galaxies, instead of
resolving single galaxies.  In this way, intensity mapping aggregates
flux from galaxies that would individually be below the detection
limit, while still resolving large-scale structure by measuring
intensity fluctuations over cosmological distances.  As a consequence,
this technique will necessarily describe galaxies in a statistical
sense through measures like the spatial power spectrum.

In this paper, we look at intensity mapping of carbon monoxide (CO), a
common tracer of molecular gas and star formation in nearby galaxies.
After \htwo{}, CO is the most abundant molecular species, tracing the metal-enriched, relatively dense ($\gtrsim 10^2$ cm$^{-3}$), cool to warm molecular ISM phase where stars form efficiently.  This physically motivates the empirical conversion between CO and \htwo{} \citep[for a review, see][]{bolatto/etal:2013}, and by extension star formation.  The CO lines themselves arise from a ``ladder'' of rotational transitions starting at millimeter wavelengths.  For this study, we focus on the ground-state CO(1-0) transition at 115.27 GHz (2.6 mm).

Other lines besides CO have also been proposed as candidate lines for
intensity mapping.  The \hi{} 21 cm line, in particular, has been
extensively studied for the purposes of tracing large-scale structure
out to $z\sim 2.5$ as well as imaging hydrogen reionization at $z
\gtrsim 6$ \citep{chang/etal:2008, morales/wyithe:2010,
  bandura/etal:2014}.  Additional potential targets include \cii{} at
158~$\mu$m \citep{gong/etal:2012, silva/etal:2014, uzgil/etal:2014},
\lya{} at 1216~\AA{} \citep{pullen/etal:2014}, and various other fine
structure lines \citep{visbal/loeb:2010}, each with its own advantages
and challenges.

Observing CO---as well as \hi{}, \cii{}, and other lines---is necessary for a census of all phases of the ISM in galaxies.
However, a practical advantage of CO is that it simultaneously emits at multiple
frequencies: because of the small excitation energies of the lowest
energy levels (5.5 and 16.6 K for the $J=1$ and 2 levels, respectively),
a galaxy is likely to have comparable emission in CO(1-0) and CO(2-1).
Observed signals in carefully chosen frequency bands can potentially
be cross-correlated to determine the contribution of the CO-only
signal.

Our focus is the ``epoch of galaxy assembly,'' at redshifts of roughly
$z \sim 2$--3.  This is a particularly interesting epoch for galaxy
formation, because it is near the peak of cosmic star formation.  Current
and near-future galaxy surveys are also expected to reach into these
redshifts, allowing the opportunity to cross-correlate the CO signal
with galaxy surveys \citep{pullen/etal:2013}.  Our current
understanding of star formation and gas content in this epoch is
incomplete, and largely limited to the bright end of the relevant
populations.  In the longer term, observations at these redshifts
could serve as a stepping stone for future CO observations that reach
into the epoch of reionization \citep{carilli:2011, gong/etal:2011,
  lidz/etal:2011}.

Previous predictions for the intensity of the CO signal vary by more
than an order of magnitude \citep[][at $z \sim 3$]{breysse/etal:2014}.
The wide range simply reflects the current scarcity of data for
typical high-redshift galaxies.  It is possible to directly simulate
these galaxies, but such simulations are expensive and still are quite
uncertain.  These uncertainties suggest a need for alternative probes
of high-redshift galaxy populations, especially over numbers and/or
volumes currently inaccessible to traditional surveys.

Given the modeling uncertainties, predictions of the expected signal
will only go so far, at least until a measurement is attempted.  Here
we also ask, what could we \emph{learn} from intensity mapping if a
measurement is made?  More precisely: given hypothetical but tractable
intensity mapping observations, what can we infer about the properties
and distribution of the underlying galaxy population?  To our
knowledge, this question has not yet been directly addressed in the
literature.  Here we put these questions in the context of CO surveys
that are now being built or planned for the near future.

The structure of this paper is as follows.  In \S\ref{sec:methods}, we
summarize our method for (1) modeling galaxy CO luminosities and
generating CO intensity maps and power spectra and (2) inferring
constraints on this model given a hypothetical observation.  In \S
\ref{sec:results}, we show how the variation of different model
parameters affects our fiducial prediction.  We then present
MCMC-inferred parameter constraints from mock intensity mapping
observations, as well as the inferred $\lir$--$\lco$ relation
and CO luminosity function.  In \S \ref{sec:discussion}, we discuss
our results, the uncertainties in our modeling, and the ability of
intensity mapping to constrain the properties of galaxy populations.  Finally,
in \S \ref{sec:summary}, we resummarize our main results and conclude.

\section{Methods}\label{sec:methods}

\begin{figure}
    \centering
    \includegraphics[width=\colwidth]{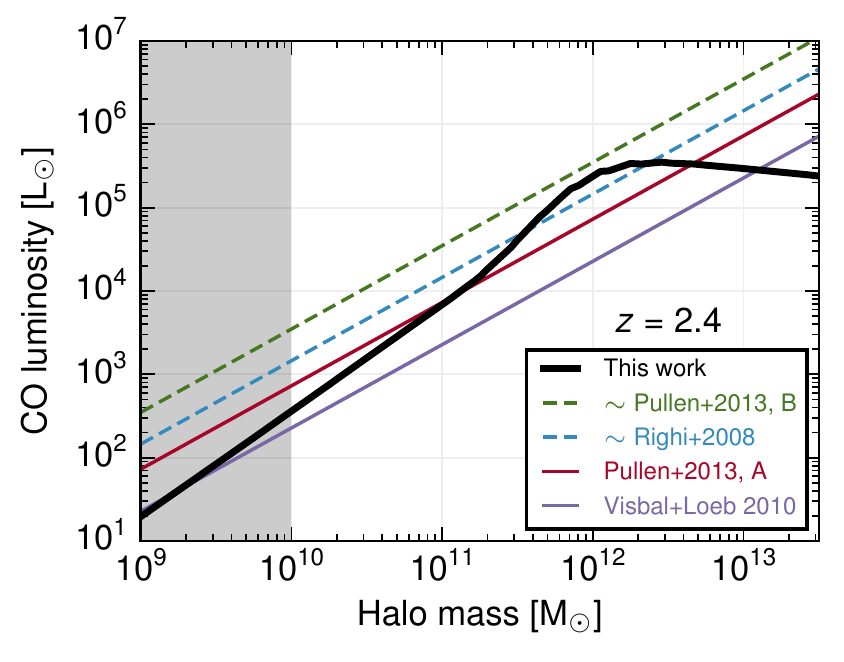}
    \caption{
        Relation between mean CO luminosity and halo mass at $z=2.4$, at the low end of the redshift range in this study.  The black solid line shows the relation used in this work.  For comparison, $\lco(\mhalo)$ from previous studies \citep{righi/etal:2008, visbal/loeb:2010, pullen/etal:2013} have also been plotted\footnote{
            \it To compare models consistently, the $\lco(\mhalo)$ relations plotted here have absorbed a ``duty-cycle'' factor, calculated in those papers as $f_{\rm duty} \approx 10^8 \text{ yr} / t_{\rm age}(z)$.  \cite{carilli:2011}, \cite{gong/etal:2011}, and \cite{lidz/etal:2011} are omitted from this plot because they focus only on reionization redshifts ($z\gtrsim 6$).
        }.
        \textit{Solid color lines} indicate the exact linear $\lco(\mhalo)$
        relations used in those studies.  \textit{Dotted color lines} indicate
        approximate linear scalings for models without a direct $\lco(\mhalo)$ relation \citep[as in][]{breysse/etal:2014}.  In
        our model, we imposed a minimum CO-luminous halo mass of
        $10^{10}\;\Msun$, excluding halos in the shaded gray range (see Appendix \ref{sec:modelvalidation} for some justification).
    } 
    \label{fig:lco_v_mass}
\end{figure}

\begin{table*}
    \centering
    \caption{Model parameters}
    \begin{tabular}{cclrll}
        Label & Link & Description & Fiducial & Prior & Details \\
        \hline
        \hline
            $\sigma_{SFR}$ & 
            Halos $\rightarrow$ SFR &
            Log-scatter in SFR & 
            0.3 &
            $\mathcal{N}(0.3, 0.1)$ and $\sigma_{\rm SFR} \geq 0$ &
            \S \ref{sec:halotosfr}
        \\
            $\log \delta_{\rm MF}$ &
            SFR $\rightarrow$ $\lir$ &
            SFR--$\lir$ scaling &            
            0.0 &
            $\mathcal{N}(0.0, 0.3)$ &
            \S \ref{sec:sfrtolir}, Eq. \ref{eq:deltamf}
        \\
            $\alpha$ &
            $\lir$ $\rightarrow$ $\lco$ &
            $\lir$--$\lco$ log-slope &
            $+1.37$ &
            $\mathcal{N}(1.17, 0.37)$ &
            \S \ref{sec:lirtolco}, Eq. \ref{eq:lirlco}
        \\
            $\beta$ &
            $\lir$ $\rightarrow$ $\lco$ &
            $\lir$--$\lco$ log-intercept &
            $-1.74$ &
            $\mathcal{N}(0.21, 3.74)$ &
            \S \ref{sec:lirtolco}, Eq. \ref{eq:lirlco}
        \\
            $\sigma_{\lco}$ & 
            $\lir$ $\rightarrow$ $\lco$ &
            Log-scatter in $\lco$ & 
            0.3 &
            $\mathcal{N}(0.3, 0.1)$ and $\sigma_{\rm \lco} \geq 0$ &
            \S \ref{sec:lirtolco}
        \\
        \hline
    \end{tabular}
    \tablecomments{
        $\mathcal{N}(\mu,\,\sigma)$ denotes a normal distribution with mean $\mu$ and variance $\sigma^2$.
    }
    \label{tab:modelparams}
\end{table*}

\begin{table*}[!htbp]
    \centering
    \caption{Instrument and Survey Parameters}
    \begin{tabular}{rlll}
        Type & Description & Pathfinder & Full \\
        \hline
        \hline
        Instrument
            & System temperature ($\tsys$)
            & 40 K          & 35 K
            \\
            & Dual-polarization feeds ($\nfeeds$)
            \footnote{Dual polarization effectively doubles the number of feeds, so in our calculations we model $\nfeeds=19$ (500) as 38 (1000) feeds.}
            & 19            & 500
            \\
            & Beam width ($\theta_{\rm FWHM}$)
            & 6\arcmin           & 3\arcmin
            \\
            & Frequency band ($\Delta\nu$)
            & 30-34 GHz         & 30-34 GHz
            \\
            & Frequency channels ($\dnu$)
            & 40 MHz        & 10 MHz
            \\
        \hline
        Survey
            & Survey area / patch ($\asurv$)
            & 2.5 deg$^2$ & 6.25 deg$^2$
            \\
            & On-sky time / patch ($\tobs$)
            \footnote{We assume that observing time is divided equally over four patches with
                $\sim$35\% observing efficiency.  Thus, 750 hr per patch equals a total of $\sim$1 year of operation.}
            & 1500 hr        & 2250 hr
            \\
        \hline
        Derived
            & CO(1-0) redshift range
            & $2.4-2.8$ & $2.4-2.8$ 
            \\
            & Instrument sensitivity
            & 1026 $\mu$K s$^{1/2}$     & 783 $\mu$K s$^{1/2}$
            \\
            & Final map sensitivity
            \footnote{This sensitivity measure is proportional to the 3D noise power spectral density (Eq. \ref{eq:pnoise}). It is independent of spectral resolution, which is important here since we are mainly interested in measuring 3D structure.  To get rms noise per channel instead, divide the value by $\sqrt{\dnu}$.}
            & 41.5 $\mu$K MHz$^{1/2}$   & 18.3 $\mu$K MHz$^{1/2}$
            \\
            & Total $P(k)$ detection significance ($\mathrm{SNR_{tot}}$)
            \footnote{See \S \ref{sec:fidpredict} for additional discussion.}
            & 7.89 $\sigma$ & 144 $\sigma$
            \\
        \hline
    \end{tabular}
    \label{tab:obsparams}
\end{table*}

\subsection{Context: Previous Studies}\label{sec:prevmodels}

Several previous studies have explored CO intensity mapping and modeled the expected signal.  To date, they include: \cite{righi/etal:2008}, \cite{visbal/loeb:2010}, \cite{gong/etal:2011}, \cite{carilli:2011}, \cite{lidz/etal:2011}, \cite{pullen/etal:2013}.

Since the models in those studies inform our own, the reader may find a brief summary useful.  Table \ref{tab:prevstudies}, located toward the end of this paper, compares the most relevant aspects of each model.  The common thread across nearly all of them \citep[except][]{gong/etal:2011} is an assumed connection between star-formation rate (SFR) and CO luminosity.  In those papers, the procedure is to first calculate SFRs and then convert to CO luminosity, though the details vary.

Around the redshifts considered in this study, \cite{breysse/etal:2014} found that the mean CO intensity predicted by some of these models differs at the extremes by more than an order of magnitude, which simply reflects the significant uncertainty in these predictions.

\subsection{Part 1: Modeling CO luminosity} \label{sec:comodel}

In our modeling, we begin with dark matter halos (tracing underlying cosmological structure) and ultimately generate simulated three-dimensional CO intensity maps.

In order, the components of this calculation are:
\begin{enumerate}
    \item Dark matter halos (\S \ref{sec:dmhalos})
    \item Star-formation rates (\S \ref{sec:halotosfr})
    \item Infrared luminosities (\S \ref{sec:sfrtolir})
    \item CO luminosities (\S \ref{sec:lirtolco})
    \item Intensity maps and power spectra (\S \ref{sec:genimapandpspec})
\end{enumerate}

Figure \ref{fig:halos2imap} illustrates the input (dark matter halos) and output (CO intensity map) of this process.  It also provides some visual intuition for the dimensions of the survey volume, as well as the cosmological structure contained within.

We discuss each of the individual steps in more detail below.  Where parameters are introduced, we state their fiducial values, as well as the priors adopted for the MCMC analysis (\S \ref{sec:mcmcmethod}).

\subsubsection{Dark Matter Halos} \label{sec:dmhalos}

We begin with dark matter halos in a ``lightcone'' volume.  To obtain
this, we use the results of a cosmological $N$-body dark matter simulation
(the \texttt{c400-2048} box\footnote{\label{fn:c125}Provided by Matthew
  Becker~(M. Becker et al.~2015, in preparation)}).  The
simulation was run
with \textsc{L-Gadget}~\citep[based on \textsc{Gadget-2},][]{springel/etal:2001, springel:2005}. 
The box has $2048^3$ particles and a side
length of 400~Mpc~$h^{-1}$, resulting in a particle mass of $5.9
\times 10^8\, M_\odot h^{-1}$. The softening length used is
5.5~kpc~$h^{-1}$, constant in comoving length.
The initial conditions are generated by
\textsc{2LPTic}\footnote{\url{http://cosmo.nyu.edu/roman/2LPT/}}~\citep{crocce/etal:2006}
at $z=99$, with the power spectrum generated by
\textsc{Camb}\footnote{\url{http://camb.info/}}.

The $\Lambda$CDM cosmological parameters of the simulation are $\Omega_m=0.286$, $\Omega_\Lambda=0.714$, $\Omega_b = 0.047$, $h=0.7$, $\sigma_8=0.82$, and $n_s=0.96$.  We assume these values throughout this paper.

Dark matter halos were identified at each simulation snapshot using
the \textsc{Rockstar} halo finder \citep{behroozi/etal:2013rockstar}.
In this work, we treated subhalos and central halos identically,
but excluding subhalos from our calculations does not significantly affect our results.

The simulation output consisted of 100 snapshots, from $z=12.33$ to
$z=0$ inclusive, with equal logarithmic spacing in $(1+z)^{-1}$,
though only the halos from six of these snapshots were ultimately
used. Lightcones were generated from these snapshots by choosing an
arbitrary $z=0$ observer origin and direction, then selecting all
halos along the line of sight within the desired survey area and
redshift range (see \S\ref{sec:obsparams}) from the appropriate
redshift snapshots.

\subsubsection{Halos and Star Formation}\label{sec:halotosfr}

To get the SFR of a dark matter halo, we use the results
of \cite{behroozi/etal:2013} and \cite{behroozi/etal:2013sfe}, which
empirically quantified the stellar mass history of dark matter halos
back to $z = 8$.  Briefly, that study constrained a parameterized stellar mass--halo mass relation by applying it to simulated halo merger trees, accounting for systematic errors and biases, and comparing derived stellar mass functions, cosmic SFRs, and specific SFRs with a comprehensive compilation of observational data.  The reader is referred to the original papers for a more detailed discussion of that study.

For our purposes, we are primarily interested in $\sfrmean (M,z)$,
their derived results for the \emph{average} halo SFR
for a given halo mass and redshift.  To simplify this calculation, we
interpolate their tabulated data \footnote{Available at the time of
  writing at \url{http://www.peterbehroozi.com/data.html}} for
$\sfrmean (M,z)$.  

The results of \cite{behroozi/etal:2013} also provide $\pm 1 \sigma$ posterior constraints on $\sfrmean$.  Note that this is distinct from halo-to-halo scatter, which we address later.  However, because these bounds span a fairly consistent log-space interval ($\sim \pm 0.15$ dex) over the relevant masses and redshifts, any variation within them is effectively a rescaling factor, which we absorb into the parameter $\deltamf$ (see below, \S \ref{sec:sfrtolir}, Eq. \ref{eq:deltamf}).

We also add halo-to-halo log-normal scatter, parameterized as $\sigma_{\rm SFR}$, since the results of \cite{behroozi/etal:2013} constrain only average halo SFRs.  This single-parameter scatter is a simple way of capturing the variation in SFR for a given halo mass.  There is evidence that ``normal'' star-forming galaxies exhibit a strong correlation with their stellar mass, with a scatter of $\sim 0.2-0.4$ dex, while starbursts may exist as a separate population with unusually high SFR \citep[e.g.][]{speagle/etal:2014, salmon/etal:2015}.  We assume the scatter in SFR given stellar mass to be reasonably similar to the scatter in SFR given halo mass.  With this in mind, we choose a fiducial value of $\sigma_{\rm SFR} = 0.3$ and a prior of $\sigma_{\rm SFR} = 0.3 \pm 0.1$.  To avoid unphysical negative scatter, we also require $\sigma_{\rm SFR} \geq 0$.

\subsubsection{Star Formation and Infrared Luminosity}\label{sec:sfrtolir}

Empirically connecting SFR to $\lco$ requires at least one intermediate, directly observable quantity, since SFRs are not directly measured.  Instead, they are inferred from photometric or spectral tracers with various underlying assumptions.  We will take this intermediate tracer to be the total infrared luminosity $\lir$ \citep[see][]{carilli:2011, lidz/etal:2011, pullen/etal:2013}, conventionally the integrated 8--1000 $\mu$m luminosity.  Thus, the model is calibrated on empirical correlations between SFR, $\lir$, and $\lco$, resting on physical assumptions about star formation, dust, and molecular gas.  Some recent work has focused on quantifying these relations at higher redshifts \citep[e.g.][for $z \lesssim 4$]{bethermin/etal:2015}, but significant uncertainties remain, especially in fainter galaxies and at higher redshifts, and part of our aim is to analyze how the signal may vary within those uncertainties.

We assume a correlation between SFR and $\lir$ \citep{kennicutt:1998} of the form
\begin{align}
    {\rm SFR} = 
    \delta_{\rm MF} \times 10^{-10} \;\; \lir
    \label{eq:deltamf}
\end{align}
where SFR is in units of $\rm \Msun \, yr^{-1}$ and $\lir$ is in units of $\Lsun$.  The normalization $\delta_{\rm MF}$ is sensitive to assumptions about the initial mass function, the duration of star formation, and dust, discussed in more detail in \S \ref{sec:modeluncertainties}.  However, it is generally calculated to be a factor of order unity \citep[$0.8 \lesssim \delta_{\rm MF} \lesssim 2$ in, e.g.][]{scoville/young:1983, thronson/telesco:1986, kennicutt:1998, barger/etal:2000, rowan-robinson:2000, omont/etal:2001}.

We adopt a \cite{chabrier:2003} initial mass function, which conventionally entails
$\delta_{\rm MF} = 1.0$ \citep[e.g.][]{magnelli/etal:2012,
  carilli/walter:2013}, as was used by \cite{behroozi/etal:2013}.

To our knowledge, there has not been a recent comprehensive study of
the expected range of $\delta_{\rm MF}$ for high-redshift galaxies,
and such a study is beyond the scope of this paper.  In order to
remain consistent with the values quoted above, we adopt a log-normal
prior of $\log\delta_{\rm MF} = 0.0 \pm 0.3$ ($\delta_{\rm MF} \approx
1.0\, {}^{+2.0}_{-0.5}$).  Note that this means the prior's $\pm 3\sigma$
interval spans nearly 2 dex.

Eq. \ref{eq:deltamf} could also be written as $SFR = \delta_{\rm MF}\delta_{\rm SB} \times 10^{-10} \lir$, where $\delta_{\rm SB}$ explicitly accounts for the fraction of $\lir$ due to starbursts, as opposed to active galactic nuclei or other sources.  For simplicity, we have absorbed $\delta_{\rm SB}$ into $\delta_{\rm MF}$, and any scatter about Eq. \ref{eq:deltamf} due to variation in galaxy type is assumed to be absorbed into $\sigma_{\rm SFR}$ (\S \ref{sec:halotosfr}).

\subsubsection{Infrared Luminosity and CO Luminosity}\label{sec:lirtolco}

To convert infrared luminosity to CO luminosity, we assume a power-law relation of the form
\begin{align}
    \label{eq:lirlco}
    \log \lir &= \alpha \log \lco^\prime + \beta
\end{align}
where $\lir$ is in units of $\Lsun$, and $\lco^\prime$ is in units of $\rm K\; km\; s^{-1}\; pc^2$ (observer units for velocity- and area-integrated brightness temperature).

For our fiducial model, we use the fit from \cite{carilli/walter:2013}, which found from a census of high-redshift galaxies $\alpha = 1.37 (\pm 0.04)$ and $\beta = -1.74 (\pm 0.40)$.

More generally, however, this relation is a rough proxy for the relation between star formation and molecular gas, and has been fit to data from high-redshift galaxies ($z \gtrsim 1$, but typically few if any galaxies beyond $z \sim 3$) in a number of studies.  These studies have found, e.g., $(\alpha,\beta)$ values of $(1.13,0.53)$, $(1.37,-1.74)$, $(1.00,2.00)$, $(1.17,0.28)$ \citep[][respectively]{daddi/etal:2010, carilli/walter:2013, greve/etal:2014, dessauges-zavadsky/etal:2014}.  These values are closely fit by the line
\begin{align}
    \alpha \approx 0.10\beta + 1.19 \label{eq:abcovar}
\end{align}
as they are generally derived from similar samples of the most luminous, and therefore detectable, high-$z$ galaxies.  We choose deliberately loose priors of $\alpha = 1.17 \pm 0.37$ and $\beta = 0.21 \pm 3.74$ by taking the mean of those four values and using their full range as the 1$\sigma$ spread.\footnote{While this paper was under review, the values of $(\alpha, \beta)$ in \cite{dessauges-zavadsky/etal:2014} were updated from $(1.19, 0.05)$ to $(1.17, 0.28)$.  We have updated text and figures in this work where appropriate.}

We also add log-normal scatter to $\lco$, fiducially parameterized by $\sigma_{\lco} = 0.3$.  This is effectively the scatter in $\lco$ for a given $\lir$ or SFR (they scale linearly and so are separated in log-space by a constant).  However, equal values of $\sigma_{\lco}$ and $\sigma_{\rm SFR}$ do not necessarily have the same effect on the signal, since SFR and $\lco$ are not required to scale linearly.  In principle, though, $\sigma_{\lco}$ and $\sigma_{\rm SFR}$ could be combined into a single scatter parameter.  We choose not to do so because empirical constraints, which inform our priors, are generally on the two separate parameters.

The prior on this parameter is $\sigma_{\rm \lco} = 0.3 \pm 0.1$, with the requirement that $\sigma_{\rm \lco} \geq 0$.  This is consistent with the results of aforementioned studies, where a scatter has been quoted.

Note that the conversion from $\lco^\prime$ (units of $\rm K\; km\; s^{-1}\; pc^2$) to $\lco$ (units of $L_\odot$) is
\begin{align}
    \lco = 4.9 \times 10^{-5}\; L_\odot
            \left(
                \vphantom{\frac{A}{A}} 
                \frac{\nu_{\rm CO,\; rest}}{\rm 115.27\; GHz} 
            \right)^3
            \left( 
                \frac{\lco^\prime}{\rm K\; km\; s^{-1}\; pc^2} 
            \right)
\end{align}
where $\nu_{\rm CO,\; rest}=115.27$ GHz is the rest-frame frequency of the CO transition.

To resummarize the model:
\begin{enumerate}
    \item Halos $\rightarrow SFR$: Get $\sfrmean (M,z)$ from the results of \cite{behroozi/etal:2013}
    \item Add log-scatter, $\sigma_{SFR}$
    \item $SFR \rightarrow \lir$: Get $\lir$ from $SFR = \delta_{\rm MF} \times 10^{-10}\, \lir$
    \item $\lir \rightarrow \lco^\prime$: Get $\lco^\prime$ from $\log\lir = \alpha\log\lco^\prime + \beta$
    \item Add log-scatter, $\sigma_{\lco}$
\end{enumerate}
with fiducial parameter values:
\begin{align*}
    &\sigma_{SFR} = 0.3 , \;
    \sigma_{\lco} = 0.3 , \\
    &\delta_{\rm MF} = 1.0 , \;
    \alpha = 1.37 , \;
    \beta = -1.74
\end{align*}

Figure \ref{fig:lco_v_mass} shows the combined result of these steps,
plotting the mean $\lco(\mhalo)$ relation from our fiducial model, as
well as the equivalent relation from previous studies.  Notably,
$\lco$ in this model is not linear in $M$, a simplifying assumption
that has been adopted in some previous studies.  This is a consequence
of our choice to model the average population including quiescent
galaxies, which results in a flatter function of mass at the high-mass
end than a model constrained using only detected star-forming
galaxies.  This can affect the shape of the power spectrum, even if
the mean CO brightness temperature is the same.

Our model may be seen as combining the two models from
\cite{pullen/etal:2013}, ``$A$'' and ``$B$'', in that we still start
with dark matter halos (their model $A$) but calculate empirically
constrained SFRs (their model $B$).  In short, we combine the
machinery for $A$ with the motivation for $B$.

\subsubsection{Generating Intensity Maps and Power Spectra}\label{sec:genimapandpspec}

Once halo CO luminosities are calculated, we generate 3D intensity maps (spectral data cubes) and power spectra.  We calculate intensities as brightness temperatures, since we expect observation and calibration to be in terms of that quantity.

To do this, the halos are binned by their positions on an RA~$\times$~Dec~$\times$~$\nuobs$ grid, where $\nuobs = \nu_{\rm CO,\; rest} / (1+z)$.  From the total CO luminosity in each cell, brightness temperatures are calculated using the Rayleigh-Jeans relation
\begin{align}
    T = \frac{c^2 I_{\rm \nu,\, obs}}{2 k_{\rm B} \nuobs^2}.
\end{align}
Here, $I_{\rm \nu,\, obs} = \lco / 4\pi D_L^2 \dnu$, where $D_L$ is
the luminosity distance to the source, and $\dnu$ is the width of the
frequency channel (spectral resolution element), specified in \S
\ref{sec:obsparams}.  This assumes that the CO line profile is
approximately a delta function.  In this study, we have not included
the effects of Doppler broadening and redshift-space distortions,
which would somewhat alter the line-of-sight signal, but they are left for
future studies.

We calculate power spectra $P(\bvec{k})$ by converting the grid to 3D
comoving units and squaring the discrete Fourier transform of the
temperature cube.  More details are provided in Appendix
\ref{sec:temp_pspec_details}.  Averaging the 3D power spectrum
$P(\bvec{k})$ in radial bins gives us the spherically averaged power
spectrum, $P(k)$.  In this paper, we plot on the quantity
$\Delta^2(k)~=~k^3 P(k) / 2\pi^2$ instead of $P(k)$ directly, mainly
to compare with previous studies.  The physical meaning of
$\Delta^2(k)$ is the amount of variance in $T$ contributed per
logarithmic interval in $k$.\footnote{ $\Delta^2(k)$ is sometimes
  called the ``dimensionless'' power spectrum if the real-space
  quantity being analyzed is unitless, e.g. number overdensity.  We
  avoid this label here because the real-space quantity of interest is
  brightness temperature, so $\Delta^2(k)$ has units of $\rm \mu K^2$.
}

We assume white (Gaussian) instrumental noise in these simulations.  As with all measurements, actual data will have additional non-ideal signal components, but we do not attempt to add these instrument-specific terms here.  By using $P(k)$ as calculated above, we implicitly assume that the CO power spectrum has been cleaned of instrument noise.  In this case, we can imagine having subtracted a constant noise power spectrum $\pnoise(k) \propto \sigma_n^2$ (Eq. \ref{eq:pnoise}) from the noisy spectrum, leaving only residual fluctuations from noise variance.  Error bars on $P(\bvec{k})$ have been calculated to account for this noise variance, as well as resolution limits and sample variance of $P(k)$.  We detail this calculation in Appendix \ref{sec:powererr}.

When showing predictions for the power spectrum, we average over 100 realizations of the same volume, where each realization is taken from a separately generated lightcone.  This averaging smoothes out fluctuations in the power spectrum that are purely sample variance.  While those fluctuations due to sample variance are important to consider, they distract from the average power spectrum predicted by the model, which is what we aim to show unless otherwise specified.

\subsection{Part 2: MCMC Inference From Mock Signal} \label{sec:mcmcmethod}

Modeling the expected signal is a useful first step.  However, given the significant empirical uncertainties of this (or any) model, there is a limit to how useful a single prediction can be.

A question of at least equal importance is how intensity mapping can inform our understanding of the underlying galaxy population.  Additionally, it is worth addressing the implications for galaxy properties \emph{if no clear CO signal} is detected by the experiment: given the uncertainties, this is a possibility.  While a non-detection might not be the most desired outcome, we wish to know whether it would be informative and therefore scientifically interesting.  To that end, we produce the following three ``mock'' power spectra:
\begin{enumerate}[leftmargin=*]
    \item $P(k)$ from our fiducial $L_{\rm CO}(M)$ model.
    \item $P(k)\approx 0$ within random fluctuations ($\sigma_P \sim \pnoise / \sqrt{\nmodes}$, the second term of Eq. \ref{eq:sigmap}).  This represents a signal consistent with a non-detection after subtracting a flat noise spectrum.
    \item $P(k)$ from the simulated data of \cite{obreschkow/etal:2009sky}, which were generated with a model different from ours.
\end{enumerate}
Combined with priors informed by existing observations, we use the
intensity mapping power spectrum to infer constraints on our model
parameter space for all three of these models.

We generate these three mock spectra and their error bars for both
proposed instruments discussed in the following section (\S
\ref{sec:obsparams}), for a total of six hypothetical signals.  The
purpose of the fiducial signal is to see how closely we can recover
the ``true'' parameter values.  The purpose of the ``zero'' power spectrum is
to investigate what constraints a non-detection places on the galaxy
population.  The purpose of the signal from \cite{obreschkow/etal:2009sky} data is to check whether our results are still reasonable when an independent model is used to generate the CO signal.

We use \texttt{emcee}\footnote{\url{http://dan.iel.fm/emcee}}, an implementation of an affine-invariant ensemble Markov chain Monte Carlo (MCMC) algorithm \citep{foreman-mackey/etal:2013, goodman/weare:2010}, to sample the posterior distributions in our model parameter space.  We provide more details, including the likelihood, in Appendix \ref{sec:appendixmcmc}.

A summary of the model parameters and their priors is provided in Table \ref{tab:modelparams}.  All parameters are sampled in linear space except for $\delta_{\rm MF}$, which is sampled in log space.

In this study we model $\sigma_{\rm SFR}$ and $\sigma_{\lco}$ as uncorrelated, so from the steps described in \S \ref{sec:comodel}, one could write the total scatter as $\sigma_{\rm tot} = (\sigma_{\rm SFR}^2/\alpha^2 + \sigma_{\lco}^2)^{1/2}$.  However, if we consider perfectly correlated or anticorrelated scatter, the total scatter would instead be $\sigma_{\rm tot} = \sigma_{\rm SFR}/\alpha + \sigma_{\lco}$ or $\sigma_{\rm tot} = | \sigma_{\rm SFR}/\alpha - \sigma_{\lco} |$.  This would change the fiducial value ($\sigma_{\rm tot} \approx 0.37$) to $\sigma_{\rm tot} \approx 0.52$ or $\sigma_{\rm tot} \approx 0.08$, respectively.  Insofar as we only care about the total scatter, these extreme cases should be adequately covered by our priors.  Strongly correlated scatter could plausibly affect the inferred values of other parameters, but we do not model this for the present study.

In this work, we have chosen to err on the side of broad priors.  A more thorough analysis of the uncertainties, systematics, and details outside the scope of this study is warranted and welcome in future papers.

\subsection{Instrument and Survey Design}\label{sec:obsparams}

To connect this work to proposed and feasible observations, we
consider two examples of possible dedicated instruments for CO
intensity mapping that can be built with current detector technology.  In our calculations, the instrument and survey
parameters determine the size and resolution of our 3D intensity map,
as well as the error bars on the power spectrum.

An essential requirement of intensity mapping is high frequency resolution. In contrast to current measurements of extragalactic backgrounds---such as the cosmic infrared background (e.g. with \emph{Spitzer} or \emph{Herschel}) or the cosmic microwave background (CMB)---which characterize excess radiation in relatively broad bands, the redshifted CO intensity from galaxies will have narrow fluctuations in frequency space, reflecting cosmological structure along the line of sight.  For reference, for CO(1-0) emission from $z=2.5$ ($\nuobs \approx 33$ GHz), a spatial depth of 1 comoving Mpc corresponds to a frequency width of 8.25 MHz, which requires high-resolution spectroscopy to resolve.

Beyond that, both instruments we consider are single-dish, ground-based telescopes covering a band of 30-34 GHz in the Ka microwave band.  These frequencies correspond to redshifted CO(1-0) emission from $z \approx 2.4$--2.8.

The first instrument is envisioned as a pathfinder experiment, with
parameters similar to the CO Mapping Array Pathfinder (COMAP), currently
under development. Its main goal would be to detect or constrain the
CO signal from galaxies at $z \sim 2$--3.  We have assumed an aperture of $\sim$6 m
with a frequency resolution of $R = \nu_{\rm obs}/\dnu
\approx 800$, with a 2.5 deg$^2$ field of view.  For the rest of the paper,
we refer to this instrument as the ``pathfinder'' experiment.

The second instrument is a more advanced experiment, which would aim to
measure the CO signal with greater sensitivity, over a wider area, and
with better resolution.  To this end, we have assumed an aperture of $\sim$12 m
aperture with $R \approx 3200$, with a 6.25 deg$^{2}$ field of view.
For the rest of the paper, we refer to this instrument as the ``full''
experiment.

The relevant parameters from the two instruments are summarized in
Table \ref{tab:obsparams}.  For the purposes of our calculations, we
assume that each instrument will pursue a survey that observes four separate
patches on the sky, a necessary survey strategy due to the limitations
of ground-based observations.  However, we can combine the power spectra from each patch, reducing
the error bars on the power spectrum (see Appendix \ref{sec:powererr})
by a factor of approximately $\sqrt{4} = 2$.

Throughout this paper, we use units of brightness temperature.  If 
desired, the unit conversion from brightness temperature $T$ to flux
density $S$ can be written as
\begin{align}
    S &= 0.108\; \mathrm{mJy\; beam^{-1}}
    \left( 
        \vphantom{\frac{A}{A}} 
        \frac{\nuobs}{32\;\mathrm{GHz}} 
    \right)^2
    \left( 
        \frac{\theta_{\rm FWHM}}{6\arcmin} 
    \right)^2
    \frac{T}{\mathrm{\mu K}}.
\end{align}

\begin{figure}
    \centering
    \includegraphics[width=\colwidth]{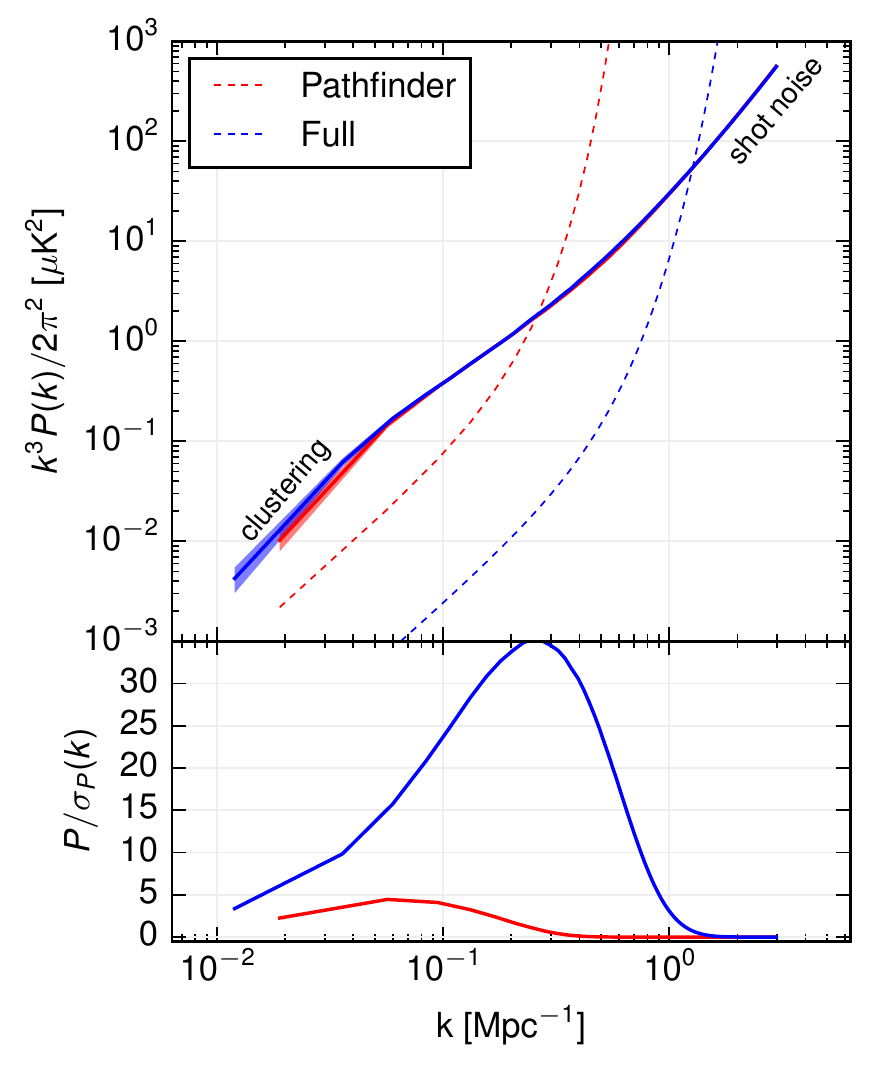}
    \caption{
        Fiducial CO power spectrum and detection significance.
        \textbf{Top}: Spherically averaged CO power spectrum from our fiducial model (\S \ref{sec:comodel}), as measured by the pathfinder (\emph{red}) and full (\emph{blue}) experiments and averaged over four separate sky patches.  \emph{Solid lines} show the measured average power spectrum.  \emph{Shaded regions} show the 1$\sigma$ uncertainty from sample variance.  \emph{Dashed lines} indicate the 1$\sigma$ limits from thermal instrument noise.  The upturn in these limits at high $k$ (small scales) arises from resolution limits (finite beam and channel width).
        \textbf{Bottom}: Detection significance of the power spectrum as a function of $k$.  The total detection significance (Eq. \ref{eq:snrtot}) is 7.89 (144) for the pathfinder (full) experiment.
    }
    \label{fig:pspec_fid}
\end{figure}

\subsubsection{Sensitivity}\label{sec:sensitivity}

We have given specific parameters for a CO intensity mapping experiment, but in this section we discuss, more broadly, the question of how we should characterize the sensitivity of an intensity mapping experiment.  There are at least a few distinct ways to address this:
\begin{itemize}[leftmargin=*]
    \item \textbf{Mean CO intensity vs instrument noise.}  
    One can compare the mean CO brightness temperature $\langle T_{\rm CO} \rangle$ to the instrument noise fluctuations $\sigma_n$ (Eq. \ref{eq:sigmanoise}), which is a useful first check \citep[e.g.][at $z\gtrsim 6$]{carilli:2011}.  However, this is not exactly the signal being measured, and it is possible to have an instrument where $\langle T_{\rm CO} \rangle < \sigma_n$ in individual map ``voxels'' (as with the pathfinder experiment in this work), and still measure cosmological intensity fluctuations on larger scales.  On those scales (across multiple voxels), random noise fluctuations should average out, while cosmological fluctuations should remain.  Moreover, an actual observation would require subtraction of continuum foregrounds, making a direct measurement of $\langle T_{\rm CO} \rangle$ (the $k=0$ mode) difficult.

    \item \textbf{Measuring spatial fluctuations in the CO signal.}  In fact, we are interested in detecting the \emph{spatial structure} of the signal.  Resolved 3D spatial structure is essential for intensity mapping.  Without it, there is no way of determining whether a signal is CO from high-redshift galaxies, which trace cosmological structure along the line of sight, or simply an unwanted systematic (e.g. synchrotron foregrounds, which are generally smooth in frequency space).  For an experiment with the goal of making a \emph{detection}, we can sum in quadrature the detection significances of each measured mode \citep[][Eq. \ref{eq:snrtot}]{pullen/etal:2013, breysse/etal:2014}.

    \item \textbf{Usefulness for inferred constraints.}  This
      requires a properly targeted range of $k$.  To measure
      cosmological large-scale structure, e.g. baryon acoustic oscillations or the linear power
      spectrum, one would like to cover large areas, focusing mainly
      on the clustering modes at low $k$.  To make inferences about
      the galaxy population, one also needs sensitivity to the shot
      noise modes at high $k$.  However, the exact transition between
      the clustering and shot noise scales depends on the details of
      the underlying galaxy population, as shown in
      \S\ref{sec:varymodel}.  One needs sensitivity to $P(k)$ in
      both regimes, either from a single experiment or from multiple
      experiments targeting different scales.  Information about the
      underlying galaxy population is contained in the relative
      strength of the power spectrum on different scales.
\end{itemize}

\begin{figure}
    \centering
    \includegraphics[width=\colwidth]{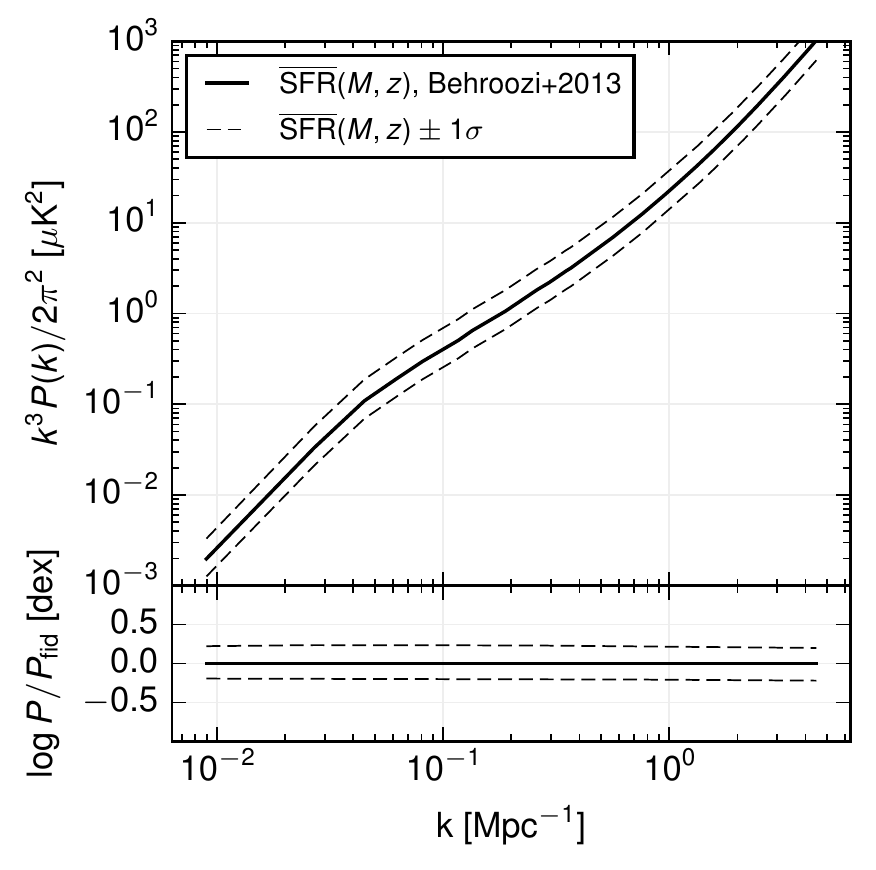}
    \caption{
        Range of power spectra spanned by $\pm 1\sigma$ uncertainties on the mean SFR$(M,z)$ from \cite{behroozi/etal:2013}.  At a fixed scale $k$, these span power spectra values of approximately $\pm 0.2$ dex.  No halo-to-halo scatter in $\lco$ has been included here (the effect of scatter is shown in Fig. \ref{fig:pspec_haloscatter}).
    }
    \label{fig:pspec_bwc13}
\end{figure}

\begin{figure*}
    \centering
    \includegraphics[height=0.385\textheight]{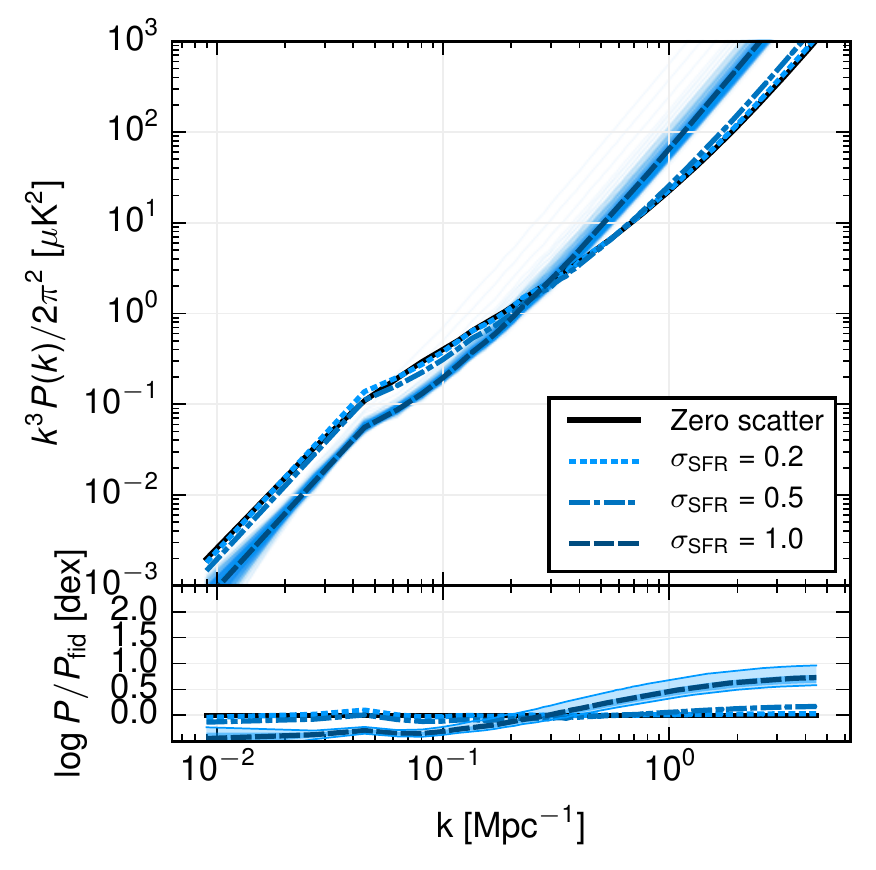}
    \includegraphics[height=0.385\textheight]{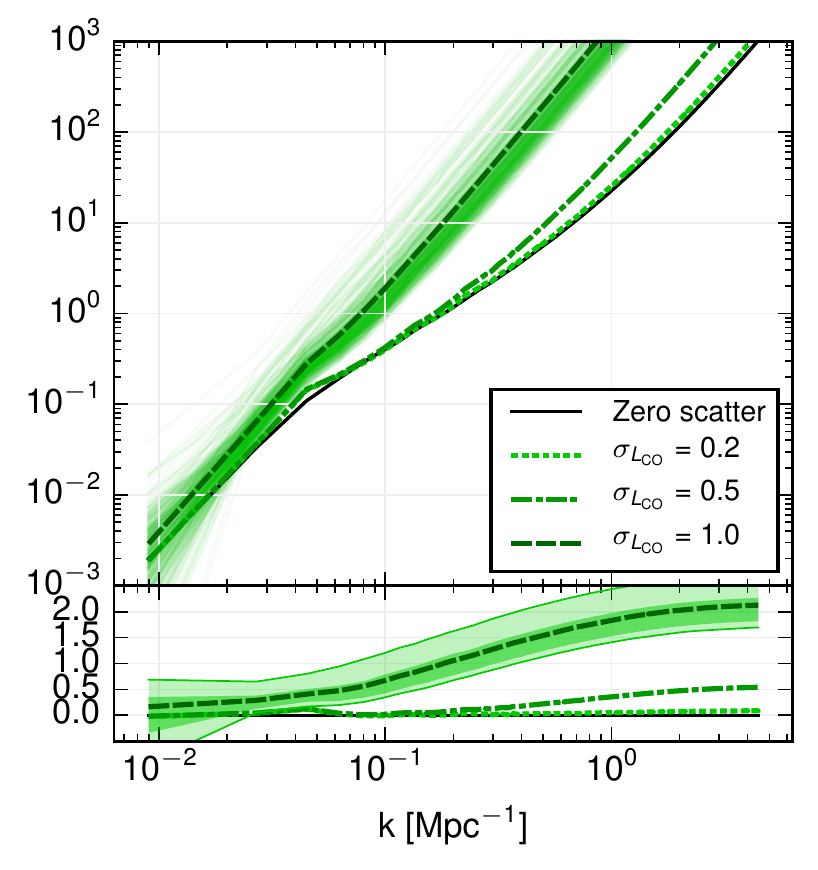}
    \caption{
        Effect on the CO power spectrum of halo-to-halo scatter as parameterized by: SFR given $\mhalo$ ($\sigma_{\rm SFR}$, \textbf{left}) or $\lco$ given SFR ($\sigma_{\lco}$, \textbf{right}).  At $\sigma_{\rm SFR}, \sigma_{\lco} \approx 1$ dex and higher, two things are evident: (1) the power spectrum begins to look like a pure shot noise spectrum, since any clustering signature is increasingly buried by the large halo-to-halo scatter, and (2) the scatter introduces significant variance into the power spectrum.  This variance is demonstrated for the $\sigma=1.0$ case by plotting 100 individual power spectra in the top plot, as well as their shaded 95\% interval in the bottom plot.  However, this scenario is probably extreme and unrealistic, because it implies a $\pm 1\sigma$ scatter over two orders of magnitude.  In this plot, the labeled lines are the mean of 100 power spectra from the exact same halos (rather than multiple realizations of the survey volume) to isolate the variance introduced by halo-to-halo scatter alone.
    }
    \label{fig:pspec_haloscatter}
\end{figure*}

\begin{figure}[!htbp]
    \centering
    \includegraphics[width=\colwidth]{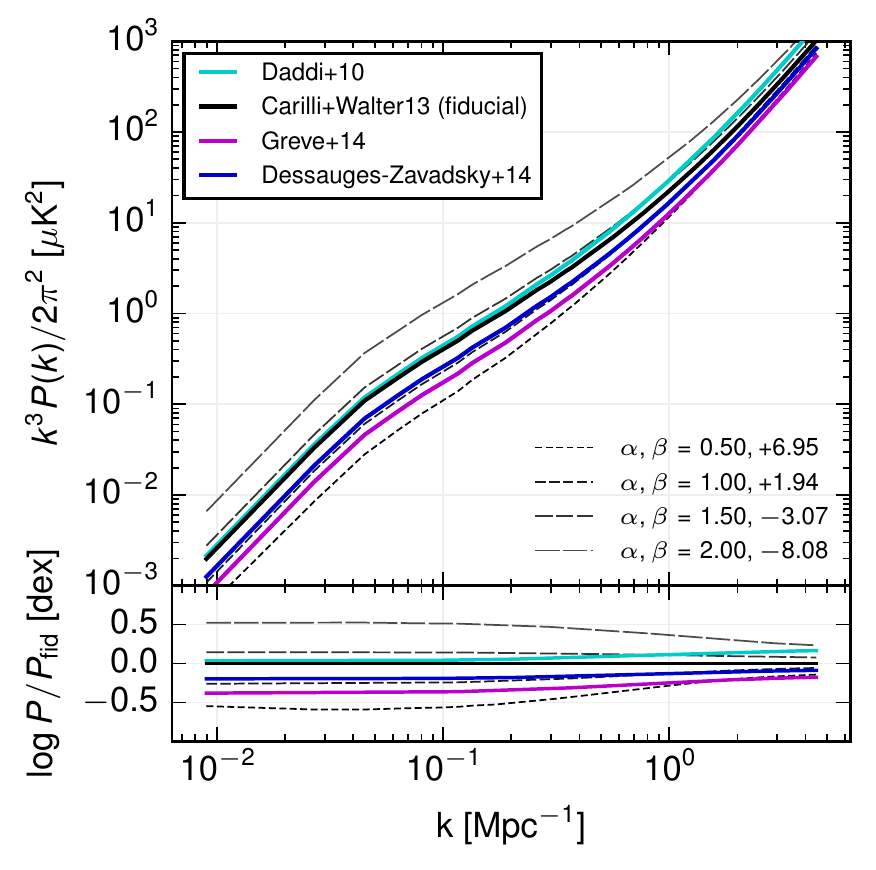}
    \caption{
        Power spectra due to varying $\lir$--$\lco$ relations.  Solid colored lines show literature values of $\alpha, \beta$ \citep{daddi/etal:2010, carilli/walter:2013, greve/etal:2014, dessauges-zavadsky/etal:2014}.  Gray dashed lines show the power spectra from $\alpha=0.5, 1.0, 1.5, 2.0$ (dashed lines) and corresponding $\beta$ assuming $\alpha$ and $\beta$ covary as in Eq. \ref{eq:abcovar}.  As $\alpha$ increases, smaller and fainter (in IR) galaxies contribute a greater proportion of the total CO luminosity, and the relative strength of the low-$k$ clustering signal increases.
    }
    \label{fig:pspec_lirlco}
\end{figure}

\begin{figure*}[!htbp]
    \centering
    \includegraphics[width=\colwidth]{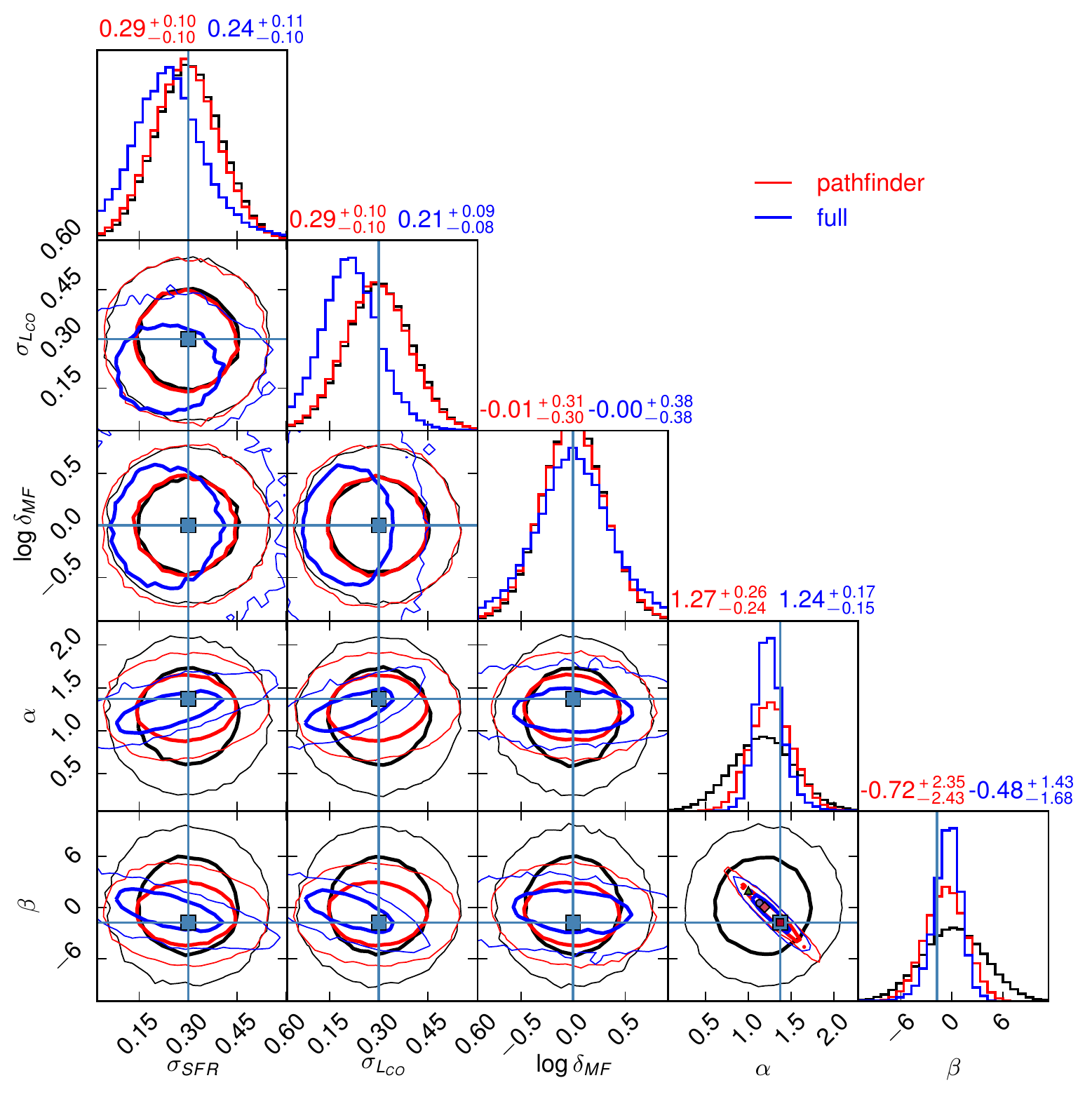}
    \includegraphics[width=\colwidth]{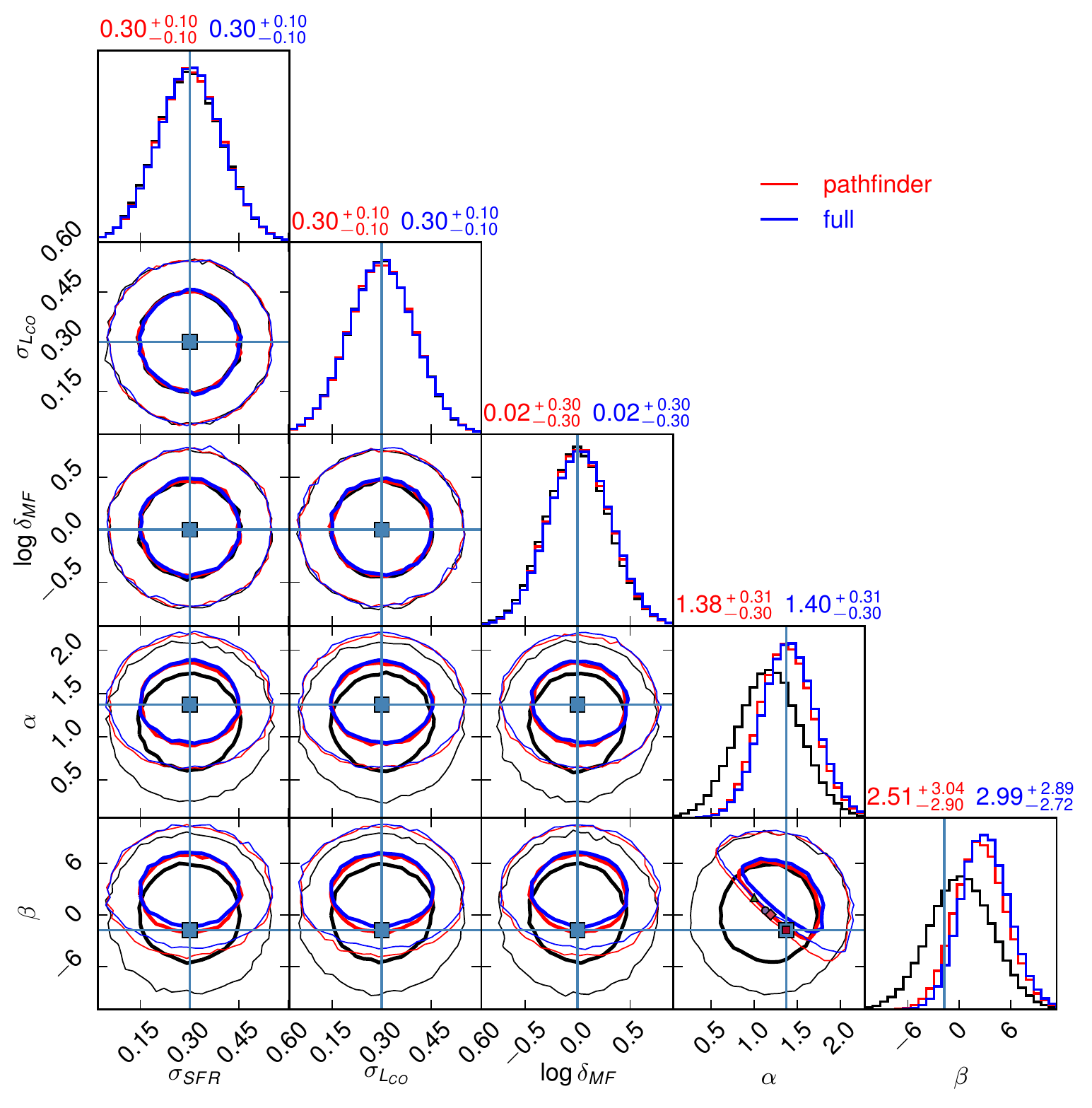}
    \caption{
        MCMC posterior distributions on the parameter space of our model.  Black contours show prior distributions.  Red contours show constraints from the pathfinder experiment, while blue contours show constraints from the full experiment.  Crosshairs indicate the values of our fiducial model.  Marked points in the $\alpha$--$\beta$ plot (bottom row, second from right) show values obtained by previous studies (see \S \ref{sec:lirtolco}).
        \textbf{Left}: Posterior distributions inferred from our fiducial signal, as observed by both the pathfinder (red) and full (blue) experiments.
        \textbf{Right}: Posterior distributions inferred from a non-detection by the same two experiments.
        }
    \label{fig:mcmc_posteriors}
\end{figure*}

\begin{figure*}[!htbp]
    \centering
    \includegraphics[width=\colwidth]{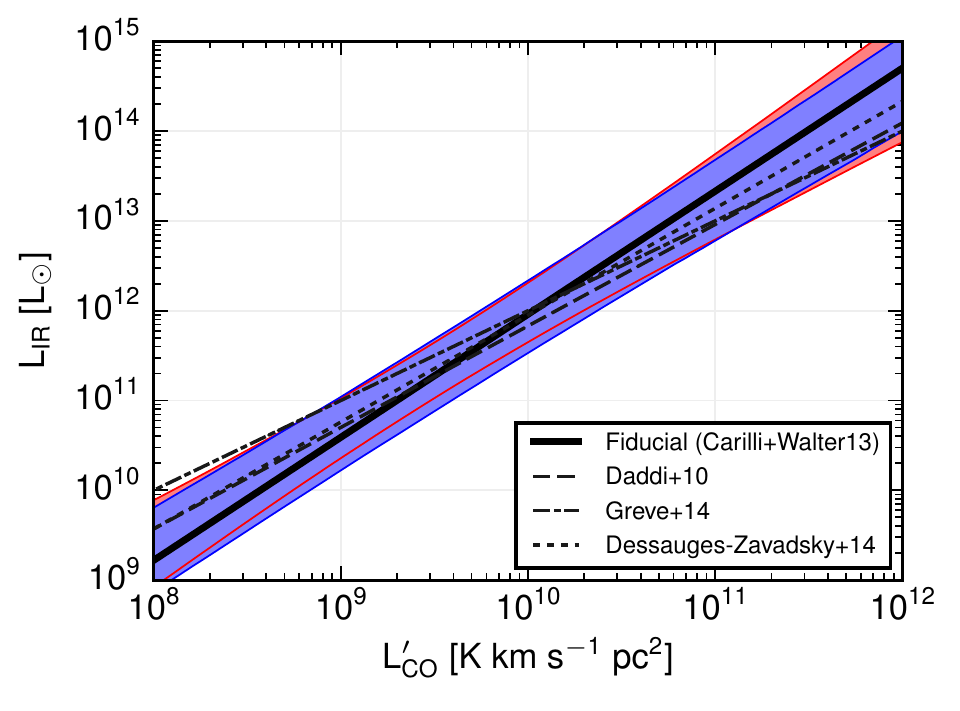}
    \includegraphics[width=\colwidth]{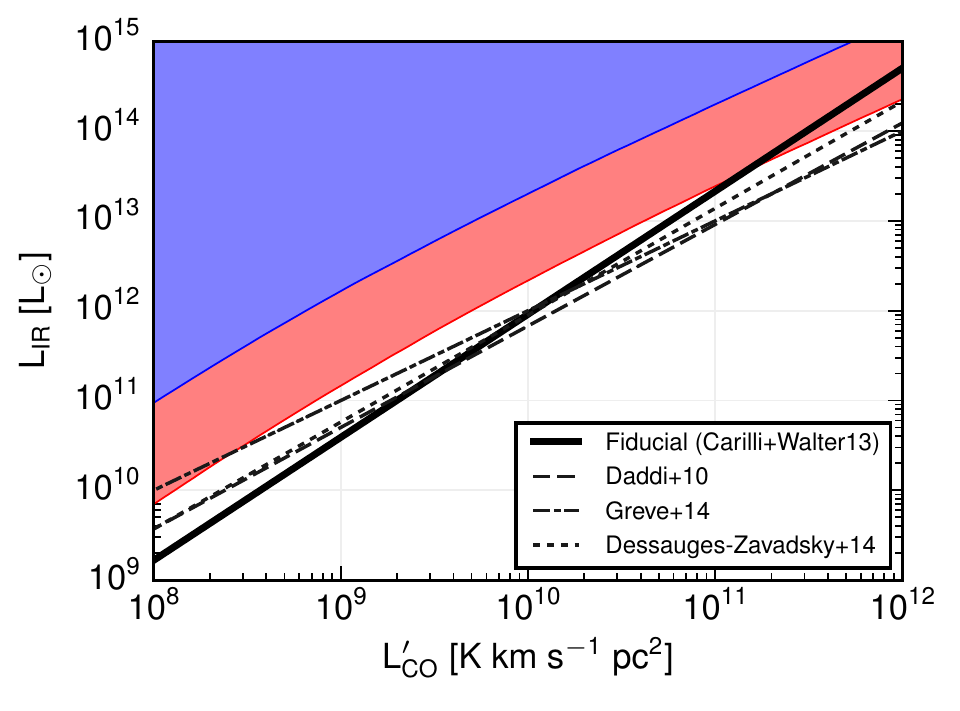}
    \\
    \includegraphics[width=\colwidth]{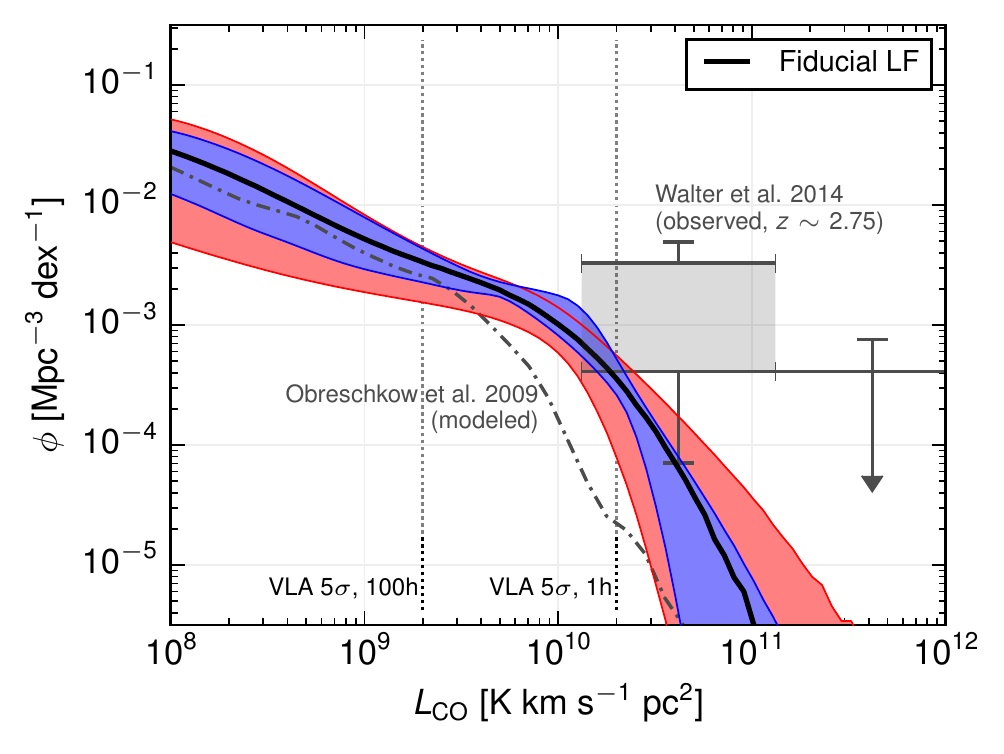}
    \includegraphics[width=\colwidth]{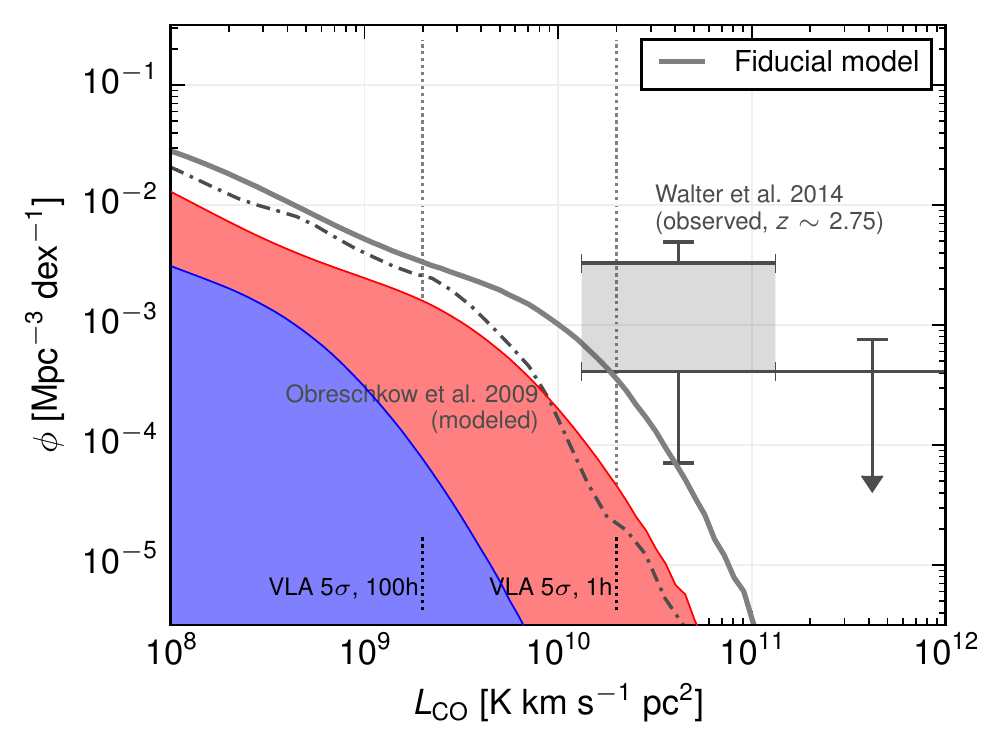}
    \caption{
        The properties of galaxy populations, as inferred by the pathfinder (red) and full (blue) experiments.
        \textbf{Top left}: Inferred 1$\sigma$ constraints on the mean $\lir$--$\lco^\prime$ relation.  Note that while the pathfinder may appear to yield comparable constraints to the full experiment, this is an artifact of our somewhat simple priors and likelihood (see \S \ref{sec:mcmcresults}).  The parameters $\alpha$ and $\beta$ are individually more constrained by the full experiment (see Fig \ref{fig:mcmc_posteriors}).  For comparison, we also plot derived fits from \cite{daddi/etal:2010}, \cite{greve/etal:2014}, and \cite{dessauges-zavadsky/etal:2014}.
        \textbf{Top right}: 95\% posterior constraints for the $\lir$--$\lco^\prime$ relation from a non-detection.
        \textbf{Bottom left}: Inferred CO luminosity functions.  Colored intervals show the median and 95\% credible interval for inferred CO luminosity functions.
        For comparison, we show observational constraints at $z\sim 2.75$ from a blind scan of the Hubble Deep Field \citep{walter/etal:2014}, as well as the model predictions of \cite{obreschkow/etal:2009} at similar redshifts.
        The vertical dashed lines marks the VLA 5$\sigma$ detection threshold assuming 1 and 100 hr of on-source time and a 100 km s$^{-1}$ channel width at 32 GHz.
        \textbf{Bottom right}: 95\% posterior constraints on the CO luminosity function from a non-detection.
    }
    \label{fig:lirlco_lf}
\end{figure*}

\begin{figure}[!htbp]
    \centering
    \includegraphics[width=\colwidth]{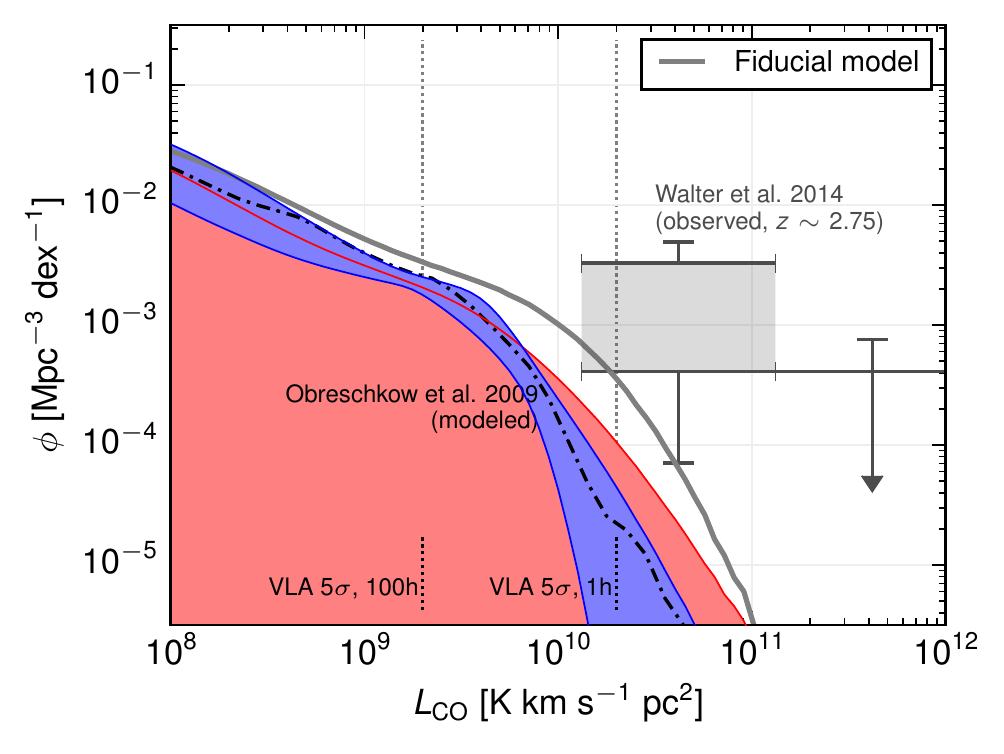}
    \caption{
        CO luminosity function from the data of \cite{obreschkow/etal:2009sky}, as inferred by the pathfinder and full experiments.  Our fiducial model has been plotted for comparison, but note that the inferred LF matches the \cite{obreschkow/etal:2009sky} curve more closely now while excluding our fiducial model (as should be expected).  For more details, see Fig. \ref{fig:lirlco_lf}.
    }
    \label{fig:lf_ob09}
\end{figure}

\section{Results}\label{sec:results}

\subsection{Fiducial Prediction and Detection Significance}\label{sec:fidpredict}

Figure \ref{fig:pspec_fid} shows the fiducial power spectra obtained from our mock intensity maps (e.g. Fig. \ref{fig:halos2imap}), as well as the mode detection significance as a function of $k$.  As has been noted by previous studies, the power spectrum may be thought of, in the context of the ``halo model'' formalism \citep{cooray/sheth:2002}, as having a clustering component (at low $k$) and a shot noise component (at high $k$).

In averaging $P(k)$, we choose bin widths of $\Delta k = 2\pi/L_{\rm min}$, where $L_{\rm min}$ is the shortest comoving dimension of the survey volume, specifically $\Delta k = 0.038$ Mpc$^{-1}$ (for the pathfinder) and 0.024 Mpc$^{-1}$ (for the full experiment).  Note that the apparent deficit in the pathfinder's low-$k$ power is because its wider, linearly spaced $k$-space bins cause the the average power in the lowest bin to appear underestimated.

Averaging the power spectrum over four identical patches, as mentioned in \ref{sec:obsparams}, we find that the maximum detection significance $P(k)/\sigma_P(k)$ for any single $k$ mode is 4.46 and 35.2 for the pathfinder and full experiments, respectively.  However, we can quantify the total detection significance of the power spectrum as \citep[e.g.][]{pullen/etal:2013, breysse/etal:2014}
\begin{align}
    \mathrm{SNR^2_{tot}} = \sum_i \left[ \frac{P(k_i) }{ \sigma_P(k_i)} \right]^2 \label{eq:snrtot}
\end{align}
where $k_i$ are the discrete values of $k$ at which the power spectrum is calculated, separated by $\Delta k$.

Assuming uncorrelated errors, we find the total signal-to-noise ratio (SNR) to be 7.89 and 144 for the pathfinder and full experiments, respectively.  However, those values are only for our specific choice of modeling and observing parameters.  In general, $\mathrm{SNR_{tot}}$ may be improved by increasing bandwidth or by decreasing noise fluctuations (the latter may be achieved by increasing observing time, increasing the number of feeds, or decreasing system temperature).

It is also possible to optimize the survey area to maximize $\mathrm{SNR_{tot}}$ \citep[see, e.g.][]{breysse/etal:2014}, though note that the exact optimal area depends on $P(k)$ and is therefore model-dependent.  When we perform this calculation, we obtain 0.6 and 9.2 deg$^2$ for the pathfinder and full experiments, respectively.  These values differ from our choices of of 2.5 and 6.25 deg$^2$, but adopting the optimal values does not yield dramatic improvements (increasing $\mathrm{SNR_{tot}}$ to 9.48 and 146).  Throughout this study, then, we maintain the pathfinder and full experiment parameters in Table \ref{tab:obsparams}.  See Appendix \ref{sec:optimizeobs} for more details and considerations.

\subsection{Varying Model Parameters}\label{sec:varymodel}

The uncertainties in $\sfrmean$ alone allow the power spectrum to span $\pm 0.2$ dex in amplitude.  This can be seen in Figure \ref{fig:pspec_bwc13}, which shows the range of power spectra expected from the the 1$\sigma$ posterior on $\sfrmean(\mhalo,z)$ \citep{behroozi/etal:2013}.  As noted, the width of this interval is fairly consistent in log space over all halo masses at the redshifts of interest ($\sim \pm 0.15$ dex).  As a result, varying $\sfrmean(\mhalo,z)$ within the posterior can be approximated as a simple rescaling, and the effect of doing so is degenerate with that of $\delta_{\rm MF}$.  In the MCMC inference procedure (\S \ref{sec:mcmcmethod}), we have absorbed the effect of this rescaling entirely into $\delta_{\rm MF}$.

Figure \ref{fig:pspec_haloscatter} shows the effect of varying halo-to-halo scatter via $\sigma_{\rm SFR}$ and $\sigma_{\lco}$.  Increasing scatter increases the relative amplitude of the shot noise component of the power spectrum, compared to the clustering ``bump'' at low $k$.  In the limit of high scatter, $P(k) \sim \mathrm{const.}$ or $\Delta^2(k) \propto k^3$.  Notably, halo-to-halo scatter introduces sample variance in the signal \emph{within the same volume}.  However, this effect only appears to be significant for very high scatter: $\sigma_{\rm SFR} \sim 1.0$ dex or $\sigma_{\lco} \sim 1.0$.  These values are not expected to be realistic since they imply a 68\% scatter of 2 dex.

We do not show the effect of varying the SFR-IR normalization $\delta_{\rm MF}$, since it is simply a scaling factor: with other parameters fixed, a higher value of $\delta_{\rm MF}$ results in fainter CO brightness.  For a given galaxy, a high value for $\delta_{\rm MF}$ ($\gg 1$) would mean a large amount of star formation per IR luminosity.  This could could indicate very low dust content or very short star formation timescales.

Figures \ref{fig:pspec_lirlco} shows the effect of varying $\alpha$
and $\beta$ in the $\lir-\lco^\prime$ relation.  Rewriting
Eq. \ref{eq:lirlco} as $\log\lco^\prime = \alpha^{-1} (\log\lir -
\beta)$ makes the effect of each parameter more apparent.
Independently increasing $\alpha$ or $\beta$ decreases $\lco^\prime$
for a given $\lir$.  However, increasing $\alpha$ also weights halos with low
$\lir$ ($\sim$ low SFR) more \emph{relative} to halos with high $\lir$
($\sim$ high SFR).  Depending on which halos are occupied by
high-SFR galaxies, this can affect the shape of the power spectrum,
particularly the clustering signal.  This is a notable difference that
can arise in predictions of power spectra if SFR is not modeled as simply
linear in $\mhalo$ (in our model, the average SFR turns downward just
above $\mhalo \sim 10^{12}\; \Msun$, a consequence of the now
well-established downturn in the ratio of stellar mass to halo mass 
at a given halo mass).

\subsection{MCMC Inference}\label{sec:mcmcresults}

Figure \ref{fig:mcmc_posteriors} shows the posterior distributions for
all model parameters, as inferred from the ``fiducial'' and
``non-detection'' power spectra.  In general, the full experiment
places stronger constraints on the model parameters than the pathfinder.

Figure \ref{fig:lirlco_lf} shows constraints on the $\lir$--$\lco^\prime$
relation and the CO luminosity function, from both the pathfinder and
full experiment observing our fiducial signal.

In the $\lir$--$\lco^\prime$ plots, we have overplotted the four
empirical fits mentioned in \S \ref{sec:lirtolco}
\citep{daddi/etal:2010, carilli/walter:2013, greve/etal:2014,
  dessauges-zavadsky/etal:2014}.  The inferred relation is consistent
with the fiducial line \citep{carilli/walter:2013} within 1$\sigma$,
and consistent with all four lines within 2$\sigma$.  The notable point
here is that while the four empirical fits were extrapolated from
a limited, bright sample of galaxies (hence their convergence around
$\sim 10^9 - 10^{10}$ K km s$^{-1}$ pc$^2$), the relation inferred from the intensity map was \emph{constrained by probing faint
  populations directly}, albeit in integrated emission.

In the plots of CO luminosity function (LF), we recover the ``true'' CO luminosity function (underlying the fiducial intensity map) to within 1$\sigma$ over nearly all luminosities.  Here, the full experiment produces tighter constraints than the pathfinder experiment, as expected.  Note that the ``true'' LF lies outside of the 1$\sigma$ interval at the bright end.  While this is not a very strong deviation, we expect that intensity mapping will not be especially sensitive to the bright end of the LF if somewhat fainter populations dominate the overall signal, as they do here.

We also show 2$\sigma$ posterior constraints on the CO luminosity function in the event that no clear signal is detected (Fig. \ref{fig:lirlco_lf}, bottom right).  While these constraints visually resemble upper limits, they are actually allowed regions and are therefore sensitive to our choice of priors.  We include further discussion of these priors in \S \ref{sec:discussconstraints}.

To compare the LF constraints with a hypothetical blind survey for CO in individual galaxies, we have marked the faintest CO luminosity that could be detected by the VLA in 1 and 100 hr of on-source time.  For the calculation, we used the VLA exposure calculator\footnote{\url{https://obs.vla.nrao.edu/ect}}, assuming an observing frequency of 32 GHz ($z \approx 2.6$), 100 km s$^{-1}$ channel, and 5$\sigma$ detection threshold.  100 hours of on-source time can probe below the ``knee'' of the luminosity function predicted by this model.  However, the relatively small field of view ($\sim 1.3 \arcmin$ primary beam at 32 GHz) means that sample variance in such a blind survey will be significant.

\subsubsection{Cross-check with Mock Data from Obreschkow et al. (2009)}\label{sec:obreschkowcompare}

The model used to generate the mock signal is the same model used in
the MCMC procedure.  It is reassuring, but perhaps unsurprising, that
we recover the ``true'' CO luminosity function when fitting this
model.  What if we use a population of CO galaxies that was generated
from an independent model applied to a different simulation?

Various studies have provided physically motivated modeling of $\lco$ at high redshift, whether through direct hydrodynamic simulation or semi-empirical modeling \citep[e.g.][]{popping/etal:2014, lagos/etal:2015}.  For this exercise, meant to provide a sanity check rather than a
prediction, we assume the underlying galaxy $\lco$'s are given by the results of
\cite{obreschkow/etal:2009, obreschkow/etal:2009sky}, who use a semi-analytic model to predict
galaxy CO luminosities from dark matter halos.  From their publicly
downloadable lightcone
data\footnote{\url{http://s-cubed.physics.ox.ac.uk/s3_sax}}, we
generate CO intensity maps and rerun the same analysis, for both the
pathfinder and full experiments.  See Appendix \ref{sec:obreschkowdata} for more details.

Figure \ref{fig:lf_ob09} compares the actual CO luminosity function in
the volume with the inferred luminosity function from intensity
mapping.  The weaker signal is consistent with non-detection for the
pathfinder.  However, it is detected by the full experiment, and within the 95\% credible intervals, the inferred LF is consistent with the "true" LF, while
our original model's LF is largely excluded.  While this is not an
exhaustive check, it is an indication that our model appears flexible
enough to accommodate a reasonable range of predicted LFs.

\begin{figure}
    \centering
    \includegraphics[width=\colwidth]{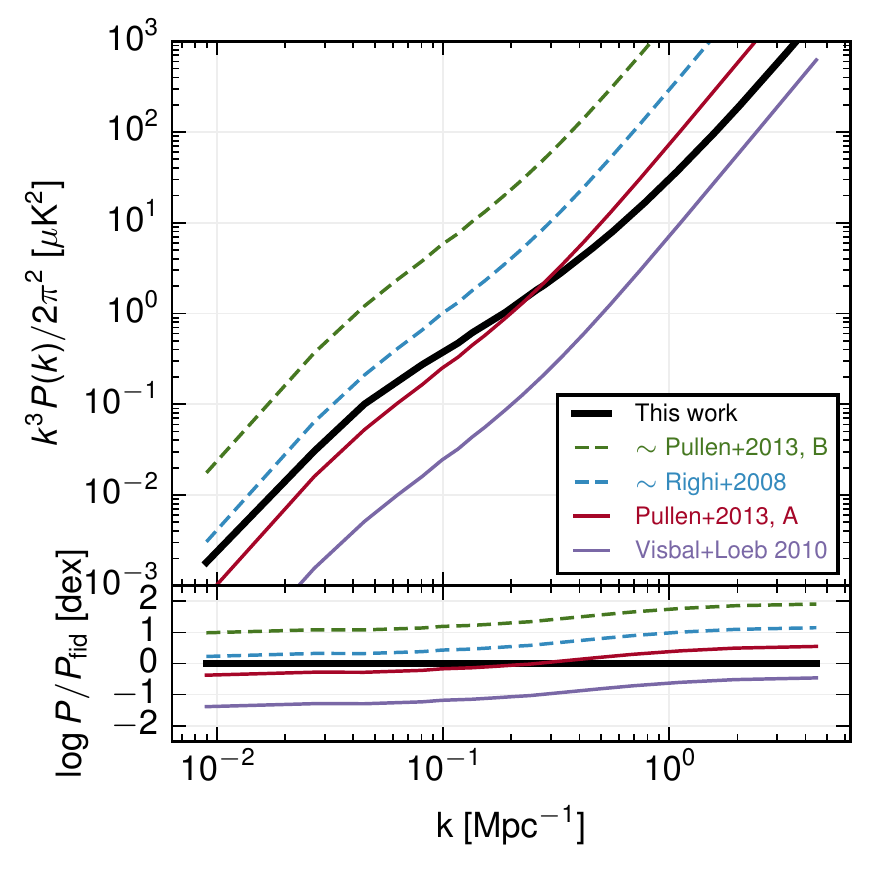}
    \caption{Comparison of the fiducial power spectrum in this work with that of previous models \citep[][see Fig. \ref{fig:lco_v_mass}]{righi/etal:2008, visbal/loeb:2010, pullen/etal:2013}, which either modeled $\lco \propto \mhalo$ or have been approximated as such here.  Our model does not have a linear $\lco(\mhalo)$, and this qualitatively affects the shape of the resulting power spectrum.  Nevertheless, our fiducial prediction appears to be consistent with the range of previous predictions.
    }
    \label{fig:pspec_compare}
\end{figure}

\section{Discussion}\label{sec:discussion}

\subsection{Relation to Previous Work}\label{sec:discuss-prev-work}

Our work here naturally extends the various studies mentioned in \S
\ref{sec:prevmodels}.  In modeling, our main improvements are (1) an
empirically constrained connection between halos and mean SFRs and (2)
an $\lir$--$\lco$ connection that is fit to a larger set of observations.
These should model more accurately the connection between underlying
large-scale structure and the predicted CO intensity map.
Qualitatively, the shape of the overall $\lco(\mhalo)$ relation is
somewhat similar to that of \cite{gong/etal:2011}, though their focus
was on $z \gtrsim 6$.

Figure \ref{fig:pspec_compare} shows the power spectrum of our
fiducial model compared to other models at similar redshifts (see
Fig. \ref{fig:lco_v_mass}).  Note that the power spectra of those models were calculated with a halo mass cutoff of $\mcomin=10^9\, \Msun$, lower than our fiducial $10^{10}\,\Msun$, to better compare with their original predictions.  However, this does not affect the main qualitative point of the plot: the nonlinear form of our $\lco(\mhalo)$
affects the shape of the power spectrum --- in this case, increasing the
relative power in clustering (low-$k$) modes.

The novel aspect of this work is an attempt to use CO intensity
mapping --- or rather, simulated observations thereof --- to make
inferences about the underlying galaxy population.  This is
particularly significant because, despite the impressive capabilities
of ALMA and VLA, it will still be difficult for them to probe
complete samples of the faint end of the CO luminosity function.
Considering that CO is the most accessible tracer of molecular gas,
relying solely on galaxy surveys may limit our ability to trace the
gas--SFR connection at high redshifts.

In our approach, we have used full $N$-body simulation, which is more computationally expensive than analytically approximating the power spectrum, even if existing halo catalogs for a
simulation are already available.  Indeed, with a simple $\lco \propto M$ model, the power spectrum
can be quickly and analytically calculated from the halo model formalism
\citep{cooray/sheth:2002}.  This is how the $\lco \propto M$ power spectra (i.e. previous models) of Figure \ref{fig:pspec_compare} were calculated.  However, directly simulating these halos is a method that more naturally
accommodates information about galaxy environment and merger histories.  For
example, merger-driven starbursts could be identified and modeled
separately from the ``normal'' star-forming population.  Modeling of
that nature is left for future studies.  Galaxy formation involves many
complex processes with diverse observational signatures, and this approach provides a framework for all of those facets to be studied more directly.

\subsection{Modeling Uncertainties}\label{sec:modeluncertainties}

One goal of intensity mapping is to probe high-redshift galaxies too
faint to be individually observed in large numbers.  By nature, then,
these are currently poorly studied populations, and any attempt to
empirically model their $\lco$'s must rely on data from (1) galaxies
thought to be low-redshift analogs, e.g. local dwarfs and spirals; (2)
observed galaxies at high redshift, which may be atypically luminous
or star-forming; or (3) galaxies that might be ``typical'' at
high redshift, but in small and incomplete samples.

Predictions about such galaxies are subject to significant
uncertainties about the formation and evolution of high-redshift
galaxies.  In particular, the nature of the ISM at high redshift is
not fully understood.  In the very early universe, one might expect
significantly lower metallicity and dust shielding in the ISM,
resulting in lower CO abundance.  However, in active star-forming regions, 
young massive stars should quickly pollute their surroundings with metals.  
These various effects have not been well quantified in typical 
high-redshift galaxies and are just beginning to be probed by individual 
galaxy observations with, e.g., ALMA and VLA.

\subsubsection{Using $\lir$ to trace SFR and $\lco$}

In our model, $\lir$ correlates with SFR (Eq. \ref{eq:deltamf}).  This
is because we need an empirical tracer of star formation to link halo
SFR with CO luminosity.  Using $\lir$ assumes that star formation occurs
in the presence of dust, and that $\lir$ is thermally
radiated from dust heated by massive, young stars.

The normalization $\delta_{\rm MF}$, as well, is sensitive to star formation
history and the initial mass function, both of which may be different at high redshift.
It should also be noted that a tracer like $\lir$ may overestimate the
instantaneous SFR if star formation occurs in short ``bursty''
episodes \citep[$\ll 100$ Myr, e.g.][]{dominguez/etal:2014} or in recently
quenched galaxies \citep{hayward/etal:2014}.

Additionally, although we have taken $\lir$ to be the total luminosity
in the $8-1000$ $\mu$m band, note that this is not always the directly
observed quantity.  $\lir$ may be converted from a measurement of flux
in a narrower IR band.  Any systematic errors introduced by this fact
will probably not be dominant, but it is something to remember going forward.

\subsubsection{CO Line Luminosities}

Existing high-redshift data for CO(1-0) luminosity are also not free
of uncertainty.  Some values for the CO(1-0) luminosity were not
directly measured, but instead inferred from higher-order transitions
that were accessible in an available observing band
\citep[e.g.][]{genzel/etal:2010, tacconi/etal:2013}.  Inferring
CO(1-0) luminosities from such observations relies on assumptions
about spectral line energy distribution (SLED), i.e. the relative
luminosities of each CO line.  In the optically thick, high-temperature limit, all lines have the same brightness temperature, or
equivalently luminosities of $L_{CO(J \rightarrow
  J-1)}~\propto~J^2~L_{\rm CO(1-0)}$, but this generally overestimates high-$J$ luminosities relative to $L_{\rm CO(1-0)}$.

While some recent work has focused on characterizing the SLED across a sample of galaxies \citep[e.g.][]{greve/etal:2014, narayanan/krumholz:2014}, the scatter is large and predicting the SLED in an individual galaxy is not straightforward.  On the other hand, it may be possible to constrain the average SLED by cross-correlating two or more intensity maps that target the same cosmological volume through different CO transitions.

\subsection{Inferred Constraints on Galaxy Populations}\label{sec:discussconstraints}

The $\lir$-$\lco^\prime$ relation (Eq. \ref{eq:lirlco}) has been fit to observations in previous literature \citep{daddi/etal:2010, dessauges-zavadsky/etal:2014, carilli/walter:2013, greve/etal:2014}.  In Figure \ref{fig:lirlco_lf}, we have plotted this relation, both as derived in the literature and as inferred from a hypothetical CO signal.  To the extent that the $\lir$--$\lco^\prime$ relation traces the star formation--molecular gas relation, it is possible for intensity mapping to characterize the latter in unobserved galaxy populations.

Our results also constrain the CO luminosity function.  Currently, this is not well characterized at the redshifts in this study because of the long integration times required to detect CO in a single galaxy.  We have overplotted the results of \cite{walter/etal:2014}, which estimated the CO luminosity function from a blind scan for CO lines in the Hubble Deep Field.

The noteworthy point here is that these constraints, especially for the faintest galaxies, arise from directly imaging their emission, albeit in aggregate.  By contrast, because galaxy surveys preferentially probe the brightest galaxies, inferences about the faintest galaxies need to be made by, e.g., extrapolation.  With intensity mapping, part of the information we receive comes directly from the (integrated) faint end of the luminosity function.

In future analyses, one can certainly consider tighter priors than we have chosen here.  In the event of a \emph{detection} of an intensity mapping signal, we expect narrower priors to yield narrower constraints on the CO luminosity function.  In the event of a \emph{non-detection}, we would actually expect narrower priors to yield less stringent constraints, i.e. a higher upper bound on the allowed luminosity function.  This is because our broad priors on $\alpha$ and $\beta$ (of the $\lir$--$\lco$ relation) actually allow a significant region of parameter space (high $\alpha$ and $\beta$) that would be in significant tension with existing $\lir-\lco$ data, even though it yields undetectable CO power spectra, consistent with the ``observed'' intensity map.\footnote{This also appears to be why the $\lir$--$\lco^\prime$ constraints in Fig. \ref{fig:lirlco_lf} do not seem to be improved by the full experiment, a somewhat counterintuitive result.}  As a result, given a mock non-detection, a region of parameter space producing unrealistically faint populations is still allowed in the posterior region.  Tighter priors would exclude that unrealistic parameter space, raising the upper bound on the allowed luminosity function.  All the same, the plots in Fig. \ref{fig:lirlco_lf} do demonstrate that a non-detection of an intensity mapping signal can constrain the faint end of the luminosity function.  Alternatively, a more complete likelihood calculation---for example, simultaneously fitting the $\lir$--$\lco^\prime$ relation to galaxy survey and intensity mapping data---should also yield stronger and more robust constraints.

One parameter we have not allowed to vary is $M_{\rm CO,\; min}$, the minimum mass of CO-luminous halos, which we fix at $10^{10}\; \Msun$.  In principle, this could also affect the signal.  However, as we note in Appendix \ref{sec:modelvalidation}, lowering the mass cutoff yields diminishing returns in our fiducial model because $\lco(M)$ decreases toward low $M$ faster than the halo number density increases.  Any variation in the signal we might get from decreasing $M_{\rm CO,\; min}$ below below $10^{10}\; \Msun$ is expected to be eclipsed by other uncertainties in the model.  However, one can certainly imagine more realistic models than those using an abrupt halo mass cutoff.  We leave investigation of these models to future studies.

A consideration in making meaningful inferences is that the spherically averaged power spectrum essentially carries two independent sources of information: (1) the amplitude of the clustering (low-$k$) signal, and (2) the amplitude of the shot noise (high-$k$) signal.  This restricts the complexity of models we can expect to meaningfully constrain.  However, there are other effects that are not modeled here but will be implemented as refinements in a future study.  These include redshift space distortions and line broadening, which will distort the signal along the line of sight.  This slightly alters the power spectrum in that direction, potentially providing additional information.  Cross-correlation could also be useful in this regard, whether with intensity maps of other lines or galaxy surveys.  The latter would include only the brightest galaxies but nevertheless trace the same cosmological structure.

It is worth noting that, although we have focused this analysis on CO, this same approach could just as well be applied to, say, \cii{} intensity mapping.  \cii{} is also thought to trace star formation in high-redshift galaxies, but through a different (ionized) phase of the ISM.


\subsection{Astrophysical Contaminants and Complexities}

In this paper, we have assumed for simplicity that the cosmological CO(1-0) map been perfectly isolated from the raw signal.  Reality is certainly more complex.

Most of the astrophysical foregrounds and backgrounds are expected to be continuum emission.  These include the CMB itself, synchrotron radiation within the Milky Way, and (redshifted) thermal dust emission from all galaxies along the line of sight.  These components are expected to be relatively smooth in frequency space, so if the instrument has enough spectral resolution, the continua can plausibly be subtracted from the signal, while the remaining fluctuations should contain the cosmological CO signal.  Note that this continuum subtraction effectively subtracts the lowest-$k$ modes ($k=0$ is the mean intensity), and so information in those modes will be reduced or removed.  If those particular modes are important for the particular analysis at hand, the survey would need to be expanded to a larger volume to recover them.

Line broadening and redshift space distortions, while not contaminants, will alter the power spectrum along the line of sight.  This and previous studies have assumed a delta-function line profile, with zero width.  However, if CO is a reasonable dynamical tracer within galaxies, then line broadening is an important consideration.  We would expect broadening to reduce small-scale (high-$k$) power along the line of sight, limiting the narrowest useful frequency channel.  For reference, a rest-frame width of 100 km s$^{-1}$ for CO(1-0) from $z\approx 2.5$ is an observed frequency width of 11 MHz (roughly the frequency resolution of the ``full'' experiment in this paper; compare 8.25 MHz $\sim$ 1 Mpc from \S \ref{sec:obsparams}).  Additionally, galaxy peculiar velocities will cause redshift space distortions, altering the apparent clustering of CO emission.  Properly accounting for these effects requires the cylindrical power spectrum $P(k_\perp, k_\parallel)$ instead of the spherical $P(k)$ in this study.

The signal may be partially contaminated by interloping non-CO spectral lines that have been redshifted into the observed frequency band.  While we expect low-$J$ CO lines to be the brightest lines in their spectral neighborhood, other lines may cumulatively add spurious fluctuations on top of the CO signal.  In particular, HCN ($\nurest = 88.63$ GHz from $z \sim 1.6-2.0$) may contribute non-negligible shot noise to the measured power spectrum.  In a 2D $C_\ell$ analysis, \cite{breysse/etal:2015} found that using pixel masking to mitigate HCN foregrounds resulted in a loss of shot noise information.  As inferences about galaxy populations are sensitive to both clustering and shot noise components, this could complicate the astrophysical interpretation of an intensity mapping measurement.  We note that, while data for the intensities of contaminant lines are currently sparse, in ALMA observations of lensed high-redshift star-forming galaxies, HCN is dimmer than CO by at least an order of magnitude \citep{spilker/etal:2014}.

Our observed band should also catch CO(2-1) from $z \approx 5.8-6.7$.  We do not expect it to dominate the CO(1-0) signal.  However, removing the CO(1-0) signal could give us a residual CO(2-1) signal, if this removal can be done robustly enough, e.g. by cross-correlating with galaxy surveys.  This would allow one band to simultaneously study two redshift ranges.

Note that we do not expect the bright 158 $\mu$m \cii{} line to be a contaminant.  For the observing band considered in this study, an interloping \cii{} line would have to originate from $z \gtrsim 55$, when the universe is $<50$ Myr old.  It is unlikely that \cii{} will be found in significant abundance at these redshifts (though it would be a remarkable discovery).

Finally, one more concern is that since the CMB would have a higher temperature at high redshift, then as an observing background it may reduce the apparent luminosity of the CO lines \citep{obreschkow/etal:2009, dacunha/etal:2013}.  The severity of this effect depends on the ISM properties of high-redshift galaxies: if the CO-luminous regions of the ISM are significantly warmer at high redshifts, the CMB is less of a concern, and there is some evidence that this may be true \citep{harris/etal:2012, daddi/etal:2014}.  However, suppose the CO(1-0) luminosity is globally reduced by a factor $0.4$, which is similar to or more pessimistic than the predictions in \cite{dacunha/etal:2013} at $z \sim 2$--3.  $P(k)$ would be reduced by a factor $0.16$, and recalculating the fiducial detection significance $\rm SNR_{tot}$ (\S \ref{sec:fidpredict}, Eq. \ref{eq:snrtot}) yields $\sim 1.67\sigma$ and $49.6\sigma$ for the pathfinder and full experiments, respectively.  While these numbers are certainly smaller, they still suggest a signal detectable by the full experiment.  Nevertheless, the effect of the CMB as an observing background remains a valid concern due to unresolved uncertainties about the high-redshift ISM.

\subsection{Prospects for Cross-correlation}
Because intensity mapping smoothes over emission from many galaxies, cross-correlation of the signal with galaxy surveys offers a promising method of isolating the signal from galaxies alone \citep[e.g.][]{pullen/etal:2013} at a specific redshift.  The underlying idea is not that CO would necessarily originate only from observed galaxies, but rather that individual galaxies and CO would trace the same large-scale structure.  As a specific example, the survey area of the pathfinder experiment is well matched to certain galaxy surveys, including the 2 deg$^2$ COSMOS field \citep{scoville/etal:2007}, which contains extensive multiwavelength data.  Cross-correlating the CO map with existing galaxies can help validate the signal and may also constrain the properties of undetected galaxies.

Within the same field, multiple intensity maps of different lines could be quite complementary.  CO is particularly well suited for cross-correlation with itself since it emits in a known ladder of emission lines, so, for example, CO(1-0) emission observed at $\nu_{\rm obs}$ should correlate with CO(2-1) emission at $2\nu_{\rm obs}$.  In addition, cross-correlation between CO and \cii{} signals from the same redshift, as well as between CO and \hi{} maps, can provide crucial information on the nature of various phases of the ISM in galaxies.

Simultaneous intensity maps of \hi{} and CO emission, in particular, would be especially useful in probing the reionization epoch \citep[e.g.][]{lidz/etal:2011}.  The former would trace the process of reionization, while the latter might trace star-forming galaxies.  If reionization is in fact driven by star-forming galaxies, cross-correlating the two maps could place interesting constraints on the morphology of the reionized intergalactic medium (via \hi{}), the nature of the reionizing sources (via CO), and the relationship between the two.

\section{Summary} \label{sec:summary}

We have performed a preliminary analysis of the ability of CO intensity mapping to probe high-redshift galaxy populations.  The following list summarizes our main results:

\begin{enumerate}
    \item Based on our fiducial assumptions, we find that \textbf{the CO(1-0) signal from $\mathbf{z \sim 2.4}$--2.8 should be detectable by a realistic intensity mapping experiment} (\S \ref{sec:fidpredict}).  Our model is consistent with a range of previous predictions, but all models carry significant uncertainties, which are not definitively resolved by current observations.

    \item \textbf{Details of the $\mathbf{\lco}$--$\mathbf{\mhalo}$ relation produce
        measurable differences in shape of the power spectrum} (\S
      \ref{sec:varymodel}).  This encodes information about the galaxy
      population in the relative strengths of clustering (low-$k$) and
      shot noise (high-$k$) modes.  This implies that sensitivity to
      fluctuations over a large dynamic range of scales provides
      better constraints on the underlying galaxy population.
        
    \item \textbf{We demonstrate the extent to which an intensity
        mapping observation can constrain the properties of galaxy populations} (\S \ref{sec:mcmcresults}), namely (1) the $\lir$--$\lco$ relation and (2) the
      CO luminosity function.  This has significance for understanding
      molecular gas and its connection to star formation in
      high-redshift galaxies, to the extent that $\lir$ traces SFR and
      $\lco$ traces molecular gas.  At high redshifts, those
      connections will require further study.
\end{enumerate}

We reiterate that intensity mapping directly probes the faint end of the luminosity function while similar inferences from galaxy surveys rely on extrapolation from observations of bright galaxies.  This suggests that intensity mapping and galaxy surveys could be complementary avenues for understanding galaxy populations at high redshifts.

In summary, we have shown that intensity mapping is a promising method of observing CO in high-redshift galaxies.  If CO traces molecular gas in these galaxies, then CO intensity mapping is a potentially informative probe of star formation and molecular gas in the high-redshift universe, particularly in galaxies that will be hard to observe individually for the forseeable future.

\acknowledgments 

T.Y.L. and R.H.W. received support from the U.S. Department of Energy under contract number DE-AC02-76SF00515 as well as from KIPAC Enterprise funds.  This work made use of computational resources at the SLAC National Accelerator Laboratory.  We thank Matthew Becker for providing access to the cosmological simulation (\texttt{c400-2048}) used in this work.  We have also used or modified the following public Python packages: Astropy\footnote{\url{http://www.astropy.org}} \citep{astropy:2013}; \texttt{triangle\_plot} \citep[for Fig. \ref{fig:mcmc_posteriors}]{triangle.py}; and \texttt{hmf} \citep[for Figs. \ref{fig:pspec_compare} and \ref{fig:tmean_v_minmass}]{murray/etal:2013}.

We are grateful to Marcelo Alvarez, Peter Behroozi, Kieran Cleary, Olivier Dor\'{e}, Andy Harris, Yashar Hezaveh, Ryan Keisler, Daniel Marrone, Gerg\"{o} Popping, Anthony Pullen, Tony Readhead, Marco Viero, and members of the COMAP collaboration for various helpful discussions and/or comments on a previous draft.  Finally, we would like to thank an anonymous referee for their constructive comments which improved this paper.

\bibliographystyle{apj}
\bibliography{comappingbib}

\begin{appendix}

\section{CO Luminosity Model: Sanity Checks} \label{sec:modelvalidation}

\begin{figure}
    \centering
    \includegraphics[width=\colwidth]{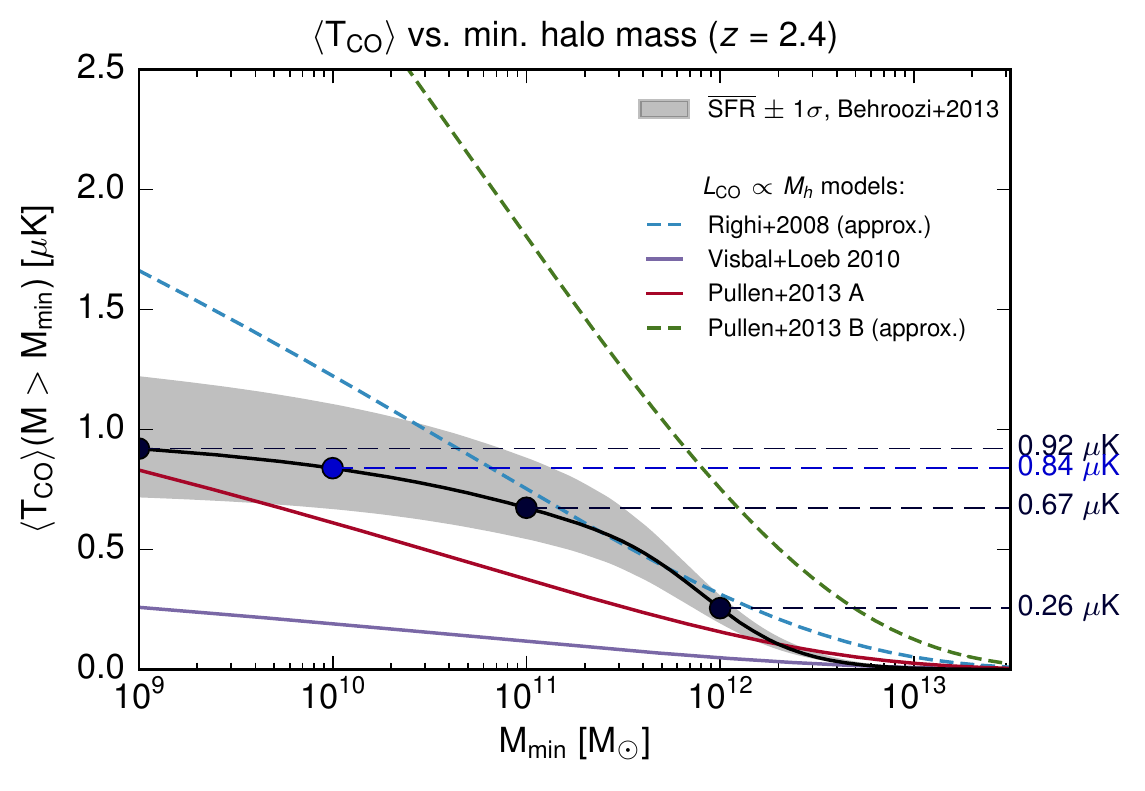}
    \caption{
        Mean CO brightness temperature as a function of minimum halo mass, at
        $z=2.4$.  As $M_{\rm min}$ increases, $\langle T_{\rm CO} \rangle$
        naturally decreases, since fewer halos contribute to total CO
        luminosity.  The shaded gray range indicates the values spanned by the
        $\langle T_{\rm CO} \rangle$ spanned by the $\pm 1\sigma$ posteriors on
        $\mathrm{SFR}(M,z)$ from \cite{behroozi/etal:2013}.  The values of $\langle T_{\rm CO} \rangle$ at $M_{\rm min}=10^9$, $10^{10}$, $10^{11}$, and $10^{12}\; \Msun$ have been labeled on the right.   Note that $\langle T_{\rm CO} \rangle$ was analytically calculated from the mass function of \cite{sheth/etal:2001}.
    }
    \label{fig:tmean_v_minmass}
\end{figure}

\begin{figure}
    \centering
    \includegraphics[width=\colwidth]{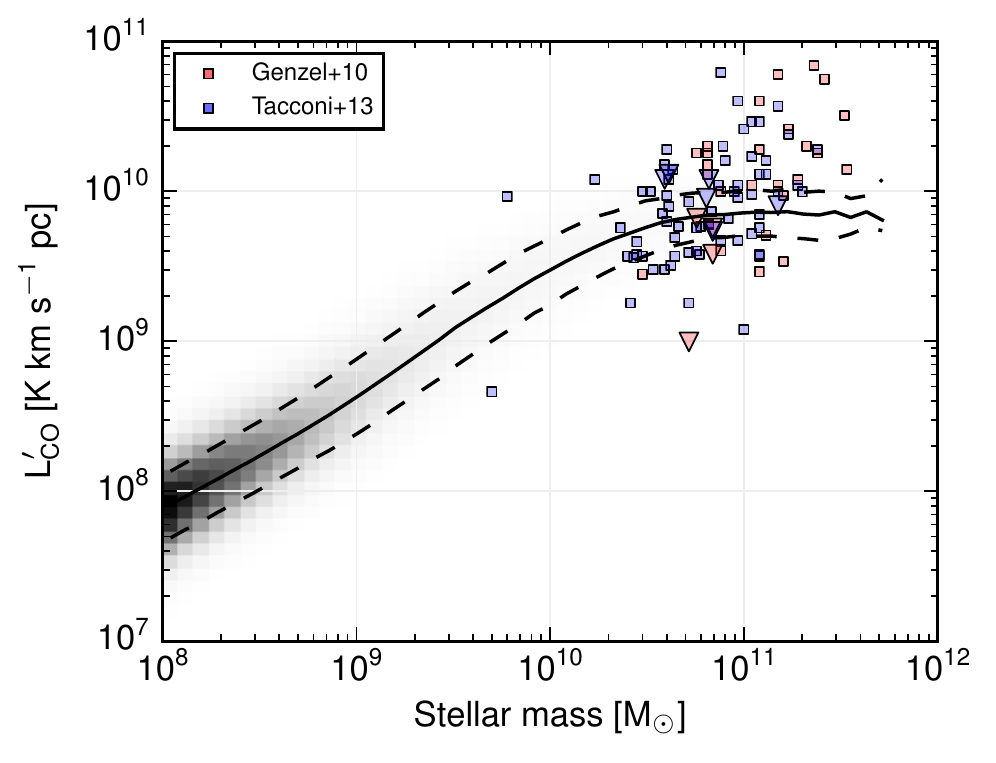}
    \caption{
        CO line luminosity vs. galaxy stellar mass.  Stellar masses were
        calculated from the stellar mass--halo mass relation of
        \cite{behroozi/etal:2013}, the same source for our SFR--halo mass
        relation.  For comparison, the observations of \cite{genzel/etal:2010} and
        \cite{tacconi/etal:2013} have been plotted: squares are measured
        values, triangles are upper limits.
        Note that the observed $\lco$ data, in their respective studies, were converted to CO(1-0) luminosities from 3-2 and 2-1 lines, assuming a certain scaling between the different line luminosities.  There is significant uncertainty in this scaling.
    }
    \label{fig:lco_v_mstar}
\end{figure}

\subsection{Minimum CO-luminous Halo Mass}

Here, we provide some justification for minimum CO-luminous halo mass of $\mcomin = 10^{10}\;\Msun$ in our model.  From a practical standpoint, this informs our choice of dark matter simulation, which is not complete below $\sim 10^{10} \Msun$.

Physically, the cutoff is motivated by the idea that smaller galaxies at higher redshifts may not be CO-luminous.  These galaxies may be less chemically evolved and/or less dusty than their low-redshift counterparts.  As a result, CO in these systems might be less abundant and/or more likely to be dissociated by ionizing radiation.  Such galaxies might not have significant CO emission despite active star formation.

A strict mass cutoff is chosen for simplicity and is unlikely to be the most realistic model.  However, the effect of loosening it is well within other model uncertainties, so the cutoff we have chosen is acceptable for the scope of this paper.

To illustrate this quantitatively, Figure \ref{fig:tmean_v_minmass}
shows the mean CO brightness temperature, $\langle \tco \rangle$, as a
function of this cutoff mass $\mcomin$.  Here, $\langle\tco\rangle$
has been obtained from the mean CO luminosity per volume, calculated as
$d\lco / dV = \int_{\mcomin}^\infty \lco(M)\, (dn/dM)\, dM$
and converting to brightness temperature, where $dn/dM$
is the analytic form of the halo mass function from
\cite{sheth/etal:2001}.  In our fiducial model at $z=2.4$, lowering
$\mcomin$ does increase the mean CO brightness temperature, but with
diminishing returns: $\lco(M)$ decreases more than the abundance of
low-mass halos increases.  To wit: in decreasing $\mcomin$ by 1 dex
from $10^{11}\;\Msun$ to $10^{10}\;\Msun$, $\langle \tco \rangle$
increases by $\sim$25\%.  In further decreasing $\mcomin$ from
$10^{10}\;\Msun$ to $10^{9}\;\Msun$, $\langle \tco \rangle$ decreases
by less than $\sim$10\% --- well within the uncertainty from $\mathrm{SFR}(M,z)$
alone --- at $M_{\rm min} = 10^{10}\;\Msun$.

Note that this particular justification is less valid if we assume
$\lco \propto M$, as in some previous models (\S
\ref{sec:prevmodels} or Table \ref{tab:prevstudies}).  Incidentally,
the fiducial models of those studies all assume $M_{\rm CO,\, min}
\leq 10^9 \Msun$, allowing more of the halo population to be CO-luminous.  Note that these models do not take into account the rapid
fall-off of the star formation efficiency below $M^*$ that is indicated
by several empirical studies (e.g. \citealt{behroozi/etal:2013} and
references therein).  For comparison, the same models from Figure
\ref{fig:lco_v_mass} have also been plotted in Figure
\ref{fig:tmean_v_minmass}.

This justification is also more uncertain at higher redshifts.  There,
for a fixed halo mass, the halo mass function grows steeper with
redshift, which may balance out the decrease in $\lco(M)$.

\subsection{Comparison with Existing $\lco$-Mass Observations}

Though sparse, measurements of $\lco$ exist for galaxies at these redshifts. Comparing the modeled $\lco(M)$ relation directly with observations is difficult, mainly because masses of dark matter halos are difficult to measure accurately.  However, measurements of \emph{stellar} mass are more commonly available from photometric data, and this is something we can compare.

Figure \ref{fig:lco_v_mstar} shows CO luminosity vs. stellar mass, both as seen in observed galaxies and as predicted by our model.  Observational data shown are from \cite{tacconi/etal:2013} and \cite{genzel/etal:2010} for $z \sim 1$--3 galaxies.  Our model does not explicitly predict stellar mass, but we have calculated it using the stellar mass--halo mass relation from \cite{behroozi/etal:2013}, which self-consistently yielded the SFR--halo mass relation used in our modeling.

The general consistency between model and data is reassuring, though
not surprising, since the $\lir$--$\lco$ relation that went into the
model was empirically fit to existing observations.  What bears the
most emphasis is the near-total lack of data points for galaxies with
low stellar mass ($\lesssim 10^{10} \Msun$), which, by number,
comprise the overwhelming majority of the galaxy population.  Note
also that our model for the average properties includes quenched
galaxies, while the data include only those galaxies detected in CO.

\section{Implementation Details}\label{sec:temp_pspec_details}

For clarity, this section describes additional details for calculating CO brightness temperature and the power spectrum.

\subsection{Halo Luminosities to 3D Intensity Map}

Once $\lco$ is calculated for all halos, they are binned on a grid according to their positions.  This grid may be defined in observational units (RA--Dec--frequency) or comoving units.  Converting between them requires us to specify a cosmology and rest-frame line frequency.  In the two dimensions across the sky, we set the grid resolution to be 10 times finer than the beam size, i.e. the voxel length in the RA and Dec directions was $\theta_{\rm FWHM}/10$.

On a comoving grid, the observed brightness temperature from a voxel of
comoving volume $\vvox$, containing total luminosity $L_{\rm CO, vox}$, emitted
at rest frequency $\nurest$, from redshift $z$ is
\begin{align*}
    T_{\rm CO,\, vox} &= \frac{c^3 (1+z)^2}{8\pi k_B \nurest^3 H(z)} \frac{L_{\rm CO, vox}}{\vvox} \\
        &= 3.1 \times 10^4 {\rm \mu K}
        \left( 1+z \right)^2
        \left( 
            \vphantom{\frac{A}{A}} 
            \frac{\nurest}{\rm GHz} 
        \right)^{-3}
        \left( \frac{H(z)}{\rm km\; s^{-1}\; Mpc^{-1}} \right)^{-1}
        \left( \frac{L_{\rm CO,\, vox} }{ {\rm L_\odot}} \right)
        \left( \frac{\vvox }{ {\rm Mpc}^3} \right)^{-1}
        .
\end{align*}

\subsection{Fourier convention and power spectrum}
We calculate the power spectrum through the Fourier transform, $F(\bvec{k})$, of the 3D CO temperature cube.  The power spectrum $P(\bvec{k})$ (specifically, power spectral density) is then
\begin{align}
    P(\bvec{k}) &= \vsurv^{-1} \left| F(\bvec{k}) \right|^2
\end{align}
where $\vsurv$ is the comoving volume of the survey.

The Fourier transform requires a choice of convention, which does not matter as
long as it is consistently applied.  However, because conventions and numerical
implementations may differ slightly, we state ours here.

For the continuous Fourier transform $F(\bvec{k})$, we use the following
(non-unitary angular frequency) convention:
\begin{align}
    \text{Forward: }
    F(\bvec{k}) &=
        \int f(\bvec{x}) \, e^{-i\bvec{k}\cdot\bvec{x}} \, d^3\bvec{x}
    \nonumber
\end{align}
with the reverse transform being
$f(\bvec{x}) = (2\pi)^{-3} \int F(\bvec{k}) \, e^{ i\bvec{k}\cdot\bvec{x}} \, d^3\bvec{k}$.
We only need the forward transform for $P(k)$, so this choice
conveniently avoids additional factors of $2\pi$.  Our discrete Fourier transform
$F_\bvec{k}$ is defined as
\begin{align}
    F_\bvec{k} &= \sum_n^{N-1} a_n \exp \left[ -2\pi i \, \frac{nk}{N} \right] ,\;\; k=0,\ldots,n-1
\end{align}
and is thus related to the continuous Fourier transform $F(\bvec{k})$ as
$F(\bvec{k}) \approx \Delta x \Delta y \Delta z \; F_{\bvec{k}}$,
where $\Delta x$, $\Delta y$, and $\Delta z$ are the comoving voxel lengths.

\subsection{MCMC Implementation} \label{sec:appendixmcmc}

We used the \emph{emcee} Python package \citep[v2.1.0,][]{foreman-mackey/etal:2013}, an implementation of an affine-invariant ensemble algorithm \citep{goodman/weare:2010}, to perform the MCMC analysis.  This approach prioritizes ease of implementation over computational speed, which we believe to be justified for the scope of this work.  To sample the five-dimensional parameter space, we used an ensemble of 200 walkers taking 2500 steps (500,000 samples in total), after a fairly generous burn-in phase of 500 steps.

At each sampled position in parameter space, we calculate a power spectrum $\pmodel(k)$ by generating a new intensity map from one of 100 realizations of the survey volume.  Because of the finite volume of these realizations, $\pmodel$ is not immune to sample variance, so it carries its own error bars, $\sigma_{\pmodel}$.

The "observed" intensity map, in reality, would appear smoothed by the telescope beam (see Figure \ref{fig:halos2imap}), i.e. convolved with a 2D Gaussian filter in each frequency channel.  This attenuates the apparent power spectrum, especially at scales smaller than the beam.  In practice, the MCMC procedure would require this convolution to be repeatedly computed, which is moderately expensive even with Fourier transforms.  Instead, we calculate the "unsmoothed" power spectrum for $\pobs$ and $\pmodel$ but increase the error bars on both by a factor $[\fres(k)]^{-1}$ (see Eq. \ref{eq:fres}).  Thus, the resolution limits are encoded in the error bars ($\sigma_{\pmodel}$ and $\sigma_{\pobs}$), rather than in $\pmodel$ and $\pobs$.

Within additive constants, the log-likelihood expression is
\begin{align}
    \ln \mathcal{L} &=
    -\frac{1}{2} \sum_k
        \left\{ 
        \frac{ \left[\pmodel(k) - \pobs(k) \right]^2}{\sigma_{\pmodel}^2(k) + \sigma_{\pobs}^2(k)}
        + \ln\left[\sigma_{\pmodel}^2(k) + \sigma_{\pobs}^2(k)\right] 
        \right\}
\end{align}
where $\sigma_{\pobs}$ is given by Eq. \ref{eq:sigmap}, and $\sigma_{\pmodel} = \pmodel / \sqrt{\nmodes} / \fres$, where $\nmodes$ and $\fres$ are functions of $k$ given by Eqs. \ref{eq:nmodes} and \ref{eq:fres}, respectively.

The 3D Fourier transform makes calculating the power spectrum the most computationally expensive part of each MCMC step, particularly as dynamic range increases (larger survey volumes and finer resolution).

\section{Error Bars on the Power Spectrum} \label{sec:powererr}

In this section, we summarize the calculation of $\sigma_P(k)$, the uncertainty in the spherically averaged power spectrum, or the error bars on $P(k)$---and by extension, $\Delta^2(k)$.  This is essential for quantifying how precisely an intensity mapping experiment can detect a signal, as well as cosmological structure within that signal.

The details that follow are previously mentioned in, e.g., \cite{lidz/etal:2011} and \cite{gong/etal:2012}.  We find that a small but non-negligible correction in the calculation is necessary (noted below).  Otherwise, we have mainly redescribed the procedure here for completeness and clarity.

This calculation accounts for three sources of uncertainty in the power spectrum:
\begin{enumerate}
    \item Sample variance.
    \vspace{-1ex}
    \item Thermal noise variance, from the instrument.
    \vspace{-1ex}
    \item Limited resolution.
\end{enumerate}
In a real experiment, there can certainly be additional systematics to consider, but this calculation is an appropriate starting point.  The equation that summarizes the calculation is
\begin{align}
    \sigma_P(k) =
        \Bigg(
            \underbrace{
                \frac{P(k)}{\sqrt{\nmodes(k)}}
            }_\text{Sample variance} +
            \underbrace{
                \frac{\pnoise(k)}{ \sqrt{\nmodes(k)} }
            }_\text{Thermal noise}
        \Bigg)
        \underbrace{
            \frac{1}{ \fres(k) }
        }_\text{Resolution}
        \label{eq:sigmap}
\end{align}
where
\begin{itemize}[leftmargin=*]
    \item $P(k)$ is the (observed) spherically averaged power spectrum, as a function of $k$

    \item $\pnoise(k)$ is the thermal noise power spectrum:
        \begin{align}
            \label{eq:pnoise}
            \pnoise(k) &= \sigma_n^2 \vvox
        \end{align}
    
    \item $\nmodes(k)$ is the number of measured modes at $k$ in a bin of width $\dk$:
        \begin{align}
            \label{eq:nmodes}
            \nmodes(k)
                &= \frac{k^2 \, \dk \, \vsurv}{4 \pi^2}
        \end{align}

    \item $\fres(k)$ is a factor that attenuates the power spectrum, arising from limited spatial resolution and mainly affecting high $k$ beyond resolvable scales:
        \begin{align}
            \fres(k) &= e^{-k^2 \sigma_\perp^2}
            \int_0^1 e^{-k^2 \left( \sigma_\parallel^2 - \sigma_\perp^2 \right) \mu^2 }\, d\mu
            \label{eq:fres}
        \end{align}
\end{itemize}
We explain Eqs. \ref{eq:pnoise}, \ref{eq:nmodes}, and \ref{eq:fres} in the subsections below.  Before we do, the following points may help to clarify notation, language, and assumptions:
\begin{itemize}[leftmargin=*]
    \item Redshift is $z$, at which $H(z)$ is the Hubble parameter, $R(z)$ is the comoving radial distance, and $D_L(z)$ is the luminosity distance.  We assume a single redshift $1+z=\nurest/\nuobs = \lambdaobs/\lambdarest$ for the volume, an acceptable approximation for the surveys being considered.

    \item Comoving Cartesian coordinates are labeled $(x_1, x_2, x_3)$ in order to avoid confusion with redshift.

    \item $x_1$ and $x_2$ are perpendicular to the line of sight (denoted $\perp$, mapping to RA/Dec).  $x_3$ is parallel to the line of sight (denoted $\parallel$, mapping to redshift or observed frequency).

    \item The instrument has system temperature $\tsys$ and number of observing feeds $\nfeeds$.  The full bandwidth of the spectrometer is $\Delta\nu$, centered on an observing frequency $\nuobs$ and divided into frequency channels of width $\dnu$.  The observing beam has a Gaussian profile, with full-width half-maximum angle $\theta_{\rm FWHM}$ ($\sigma_{\rm beam} = \theta_{\rm FWHM}/\sqrt{8\ln 2}$).  The total observing time on the survey area is $\tobs$.

    \item The smallest resolvable 3D volume element is approximated as a cubic ``voxel,'' while the sky projection of all voxels along a line of sight is a ``pixel'' (subscripted ``vox'' and ``pix'').  $\vvox$ is the comoving volume of a voxel, while $\apix$ is the solid angle on the sky subtended by a pixel.  The expressions for each are
        \begin{align}
            \apix = \sigma_{\rm beam}^2 
            \qquad
            \textrm{and}
            \qquad
            \vvox = \left[ R(z) \sigma_{\rm beam} \right]^2
                \left[ \frac{c}{H(z)} \frac{\dnu (1+z)^2}{\nurest} \right].
        \end{align}

    \item $\vsurv \approx L_1 L_2 L_3$ is the comoving volume of the survey, while $\asurv$ is the solid-angle sky coverage of the survey.
\end{itemize}

\subsection{Noise Power Spectrum [$\pnoise(k)$, Eq. \ref{eq:pnoise}]}

Consider any voxel in the observed 3D temperature cube, covering a sky area $\apix$ and frequency bin $\delta\nu$ as previously described.  We assume that the random thermal noise fluctuations, i.e. a white Gaussian noise spectrum.  By the radiometer equation, the rms temperature fluctuation per voxel is
\begin{align}
    \sigma_n = \frac{\tsys}{\sqrt{\nfeeds\, \dnu\, \tau_{\rm pix}}}
    \label{eq:sigmanoise}
\end{align}
where $\tau_{\rm pix} = \tobs (\apix/\asurv)$ is the observing time per sky pixel.

The averaged power spectrum of this noise is a constant:
\begin{align}
    \nonumber
    \pnoise(k) &= \sigma_n^2 \vvox .
\end{align}

\subsection{Number of Modes [$\nmodes(k)$, Eq. \ref{eq:nmodes}]}

Because $P(k)$ is
obtained from averaging $P(\bvec{k})$ over a finite number of modes (points in
$k$-space), we need to know the number of independently measured modes used to compute the
average.  We label this number $\nmodes$ and thus a factor $1/\sqrt{\nmodes}$
ultimately enters into the calculation of $\sigma_P(k)$.

Written explicitly, $\nmodes$ depends on both the scale of the fluctuation,
$k$, and the choice of bin width, $\dk$.  Consider a radial $k$-space bin (i.e.
spherical shell) between $k$ and $k+\dk$.  Assuming $\dk$ is chosen, the number of modes within this
shell is
\begin{align}
    \nmodes(k) &= n(k) \cdot 4\pi k^2 \dk \label{eq:nmodesdef}
\end{align}
where $n(k)$ is the number density of modes in $k$-space.  Recalling the
similar ``density of states'' calculation in statistical mechanics, $n(k)$ is
\begin{align}
    n(k) &= \left(  \frac{2\pi}{L_1} \frac{2\pi}{L_2} \frac{2\pi}{L_3} \right)^{-1}
    = \frac{\vsurv}{8\pi^3}.
    \label{eq:densitystates}
\end{align}
However, the power spectrum of a real-valued function is symmetric about
$\bvec{k}=0$, so only half of the modes contain independent
information.  This introduces a factor $1/2$ into the final expression, which is (from combining Eq. \ref{eq:nmodesdef} and \ref{eq:densitystates}):
\begin{align*}
    \nmodes(k, \dk)
        &= \frac{k^2 \, \dk \, \vsurv}{4\pi^2} .
\end{align*}

\subsection{Resolution Limits [$\fres(k)$, Eq. \ref{eq:fres}]}

Finite spatial
resolution means that information from high-$k$ modes is lost, due to smoothing
of features smaller than the beam or channel width.  The basic reasoning here is
that if the smoothed power spectrum $P_{\rm sm}(k)$ results from attenuating the ``intrinsic'' power
spectrum $P(k)$ by a factor $\fres(k)$ after smoothing, the uncertainty
should scale as $\sigma_P(k) \propto P(k)/P_{\rm sm}(k) = 1/\fres(k)$.

With arbitrarily fine resolution, the temperature field is $T(x_1,x_2,x_3)$, which
has a power spectrum $P(k)$.  In reality, we measure a smoothed field, $T_{\rm
sm}(x_1,x_2,x_3)$, which is the original field convolved with a Gaussian in each
direction:
\begin{align}
    T_{\rm sm}(\bvec{x}) &= T(x_1,x_2,x_3) * G(x_1|\sigma_1) * G(x_2|\sigma_2) * G(x_3|\sigma_3)
\end{align}
where $G(x_1|\sigma_1) = (2\pi\sigma_1)^{-1/2} \exp[-x_1^2/(2\sigma_1^2)]$
(similarly for $x_2$ and $x_3$) is a normalized Gaussian function, and ``$*$''
indicates convolution.  In reality, the beam profile may not be perfectly
Gaussian, and the frequency channels would be discrete bins.  However, the
approximation will suffice here.

Angular resolution determines $\sigma_1$ and $\sigma_2$, while frequency
resolution determines $\sigma_3$.  Specifically,
\begin{align}
    \sigma_1 = \sigma_2 \equiv \sigma_\perp = R(z)\, \sigma_{\rm beam}
    \qquad
    \textrm{and}
    \qquad
    \sigma_3 \equiv \sigma_\parallel = \frac{c}{H(z)} \frac{\dnu (1+z)}{\nuobs} .
\end{align}
Since a convolution in real space is simple multiplication in Fourier space,
the Fourier transform and power spectrum are
\begin{align}
    \widetilde{T}_{\rm sm}(\bvec{k}) =
        \widetilde{T}(k_1,k_2,k_3)\,
        e^{- (k_1^2/\xi_1^2 + k_2^2/\xi_2^2 + k_3^2/\xi_3^2)/2} 
    \qquad
    \textrm{and}
    \qquad
    P_{\rm sm}(\bvec{k}) =
        P(k_1,k_2,k_3)
        e^{- (k_1^2/\xi_1^2 + k_2^2/\xi_2^2 + k_3^2/\xi_3^2)}
\end{align}
where $\xi_1 = \xi_2 = 1/\sigma_\perp$ and $\xi_3 = 1/\sigma_\parallel$.  The Fourier convention used here ensures the transformed Gaussian carries no normalization constant.

We note a small but consequential correction: \cite{lidz/etal:2011} and some following studies state $\xi_\perp = 2\pi/[R(z)\sigma_{\rm beam}]$, but we find from the calculation above that it should actually be $\xi_\perp = 1/\sigma_\perp = 1/[R(z)\sigma_{\rm beam}]$ (without the factor of $2\pi$).

To get the spherically averaged power spectrum, define the variable
$\mu=\cos\vartheta$, where $\vartheta$ is the spherical polar angle in
$k$-space.  Then we define components of $k$ parallel and perpendicular to the
$z$ direction, $k_\parallel$ and $k_\perp$, as
\begin{align}
    k_\parallel = k_3 = k\mu 
    \qquad
    \textrm{and}
    \qquad
    k_\perp = \sqrt{\smash[b]{k^2 - k_\parallel^2}} = k\sqrt{1-\mu^2}
\end{align}
and the expression for the power spectrum becomes
\begin{align}
    \nonumber
    P_{\rm sm}(\bvec{k})
    &= P(k_\perp, k_\parallel)\,
        e^{-(\sigma_\perp^2 k_\perp^2 + \sigma_\parallel^2 k_\parallel^2)} \\
    &= P(k, \mu)\,
        e^{-k^2 \sigma_\perp^2} e^{-\mu^2 k^2 \left( \sigma_\parallel^2 - \sigma_\perp^2 \right) }.
\end{align}
To get the spherically averaged power spectrum, this needs to averaged over $0
< \mu < 1$ (the upper half-space, since only half the modes are independent) at
fixed $k$.  In this paper, we assume $P(k,\mu)$ is isotropic, so only the
exponential needs to be averaged:
\begin{align}
    P_{\rm sm}(k)
    &= P(k) \;
        \underbrace{
            e^{-k^2 \sigma_\perp^2}
            \int_0^1 e^{-k^2 \left( \sigma_\parallel^2 - \sigma_\perp^2 \right) \mu^2 }\, d\mu
        }_{\fres(k)}.
\end{align}
The expression after $P(k)$ is $\fres(k)$, which can be integrated numerically.

\section{Optimizing Observing Parameters}\label{sec:optimizeobs}

The expression for $\sigma_P(k)$ allows us to draw some conclusions about the effect of varying survey and instrument parameters.  See \cite{breysse/etal:2014} for a similar analysis with $C_\ell$ coefficients, i.e. a 2D power spectrum.

We ignore the factor $\fres(k)$ for this analysis, since its main effect is to
limit the maximum $k$ (smallest scale) at which at which meaningful
measurements can be made.  Certainly, it will also slightly suppress
power at low $k$ (large scales), but this should not be a dramatic effect.
Then, the expression for $\sigma_P(k)$, showing explicit dependences on
instrument and survey parameters, is ultimately
\begin{align}
    \sigma_P(k) = 
        \frac{2\pi}{k \sqrt{\dk \; \Delta\nu}}
        &\left[
            \frac{1}{\sqrt{\lambdarest\asurv}} \frac{\sqrt{H(z)}}{D_L(z)} \, P(k)
            + \frac{\tsys^2}{\nfeeds\tobs} \sqrt{\lambdarest\asurv} \frac{D_L(z)}{\sqrt{H(z)}}
        \right].
        \label{eq:sigmap_obsparams}
\end{align}
where $D_L(z) = (1+z)R(z)$ is the luminosity distance.  Note that redshift $z$ is directly determined by $\nu_{\rm obs}$.

Written in this form, it is easier to see how instrument and survey parameters ultimately affect sensitivity.  At a given scale $k$, the following adjustments will decrease $\sigma_P(k)$:
\begin{itemize}[leftmargin=*]
    \item {\bf Increasing bandwidth}: $\sigma_P \propto 1/\sqrt{\Delta\nu}$.  This expands the survey volume, increasing the density of modes in the $k_\parallel$ direction and thus $\nmodes$.  However, the observed redshift range will eventually grow large enough that cosmic evolution across the volume becomes important.
    \item {\bf Decrease instrument noise}: if $\sigma_P$ is noise-dominated, $\sigma_P \propto \tsys^2/\nfeeds\tobs$.  In that case, decreasing $\tsys$, increasing $\nfeeds$, or increasing $\tobs$ will all serve to increase sensitivity.  Eventually, though, decreasing system noise yields diminishing returns as sample variance becomes the dominant component of $\sigma_P$.
    \item {\bf Optimize survey area}.  For a given mode $k$, the optimal survey area results in equal sample variance ($\propto \asurv{}^{-1/2}$) and noise variance ($\propto \asurv{}^{1/2}$).  In steradians, this is
        \begin{align}
            \Omega_{\rm surv,\, opt} &= 
            \frac{1}{\lambdarest} \,
            \underbrace{\frac{\nfeeds \tobs}{\tsys^2}}_\text{From noise} \,
            \underbrace{\frac{H(z)}{[D_L(z)]^2}}_\text{From $\nuobs$} \,
            P(k).
        \end{align}
        Because $\Omega_{\rm surv,\, opt} \propto P(k)$, the predicted optimal area is model-dependent.
        
        As noted in \S \ref{sec:fidpredict}, we can maximize the total detection significance, $\mathrm{SNR_{tot}}$ (Eq. \ref{eq:snrtot}), by optimizing the area.  For the observing parameters in this study, we obtain 0.6 and 9.2 deg$^2$ for the pathfinder and full experiments, respectively.  Certainly, these optimal areas differ from our adopted survey areas (2.5 and 6.25 deg$^2$), but the improvement in $\mathrm{SNR_{tot}}$ is not dramatic (7.89 to 9.48 for pathfinder; 144 to 146 for full).
                
        However, in practice, there are other factors to consider.  The minimum survey area cannot be arbitrarily small, because the arrangement of multiple feed horns on the instrument images a pattern of multiple beams on the sky.  This pattern must be scanned across the sky to adequately cover the survey area, and limitations on the scanning strategy effectively set the minimum size of the survey area.  This motivated our estimate of 2.5 and 6.25 deg$^2$ as the survey areas for the pathfinder and full experiments, respectively.
        
        The maximum survey area can be as large as the entire sky, but increasing total area would reduce the amount of integration time \emph{per unit area}.  This can decrease sensitivity to high-$k$ (shot noise) modes, which contain important astrophysical information.
\end{itemize}

It is worth noting that the following parameters do \emph{not}, for the simple assumptions above, affect $\sigma_P$ (neglecting secondary effects such as redshift evolution across the volume and the curvature of the boundary surface, which we expect to be subdominant for the volumes being considered):
\begin{itemize}[leftmargin=*]
    \item {\bf Resolution}.  Assuming the spatial scales being probed are much larger than the resolution limit, refining the resolution does not decrease $\sigma_P$ significantly.
    \item {\bf Survey area aspect ratio}.  For a fixed rectangular survey area, its relative dimensions have no effect.  As the length grows, the width shrinks.  In $k$-space, the density of modes becomes sparser in one direction but greater in the other, and $\nmodes$ does not change.
\end{itemize}

\section{Obtaining Data from Obreschkow et al. (2009)}\label{sec:obreschkowdata}

For the results of \S \ref{sec:obreschkowcompare}, we downloaded the lightcone data of \cite{obreschkow/etal:2009sky} from their publicly accessible database.  We selected all galaxies within a central 6.25 deg$^2$ area, located at comoving distances between 5770 and 6285 Mpc, which encloses the redshift range $2.4 < z < 2.8$ in our assumed cosmology.  From each simulated galaxy, we obtained four quantities: RA, Dec, comoving distance, and CO(1-0) flux.  On the query form page\footnote{\url{http://s-cubed.physics.ox.ac.uk/queries/new?sim=s3_sax}}, the exact SQL query given was
\begin{lstlisting}[
           language=SQL,
           showspaces=false,
           basicstyle=\ttfamily,
           breaklines=true
        ]
    select ra, decl, cointflux_1, distance 
    from   galaxies_line 
    where  ra between -1.25 and 1.25 
           and decl between -1.25 and 1.25 
           and distance between 5770 and 6285
\end{lstlisting}

Since CO(1-0) flux was provided as an obesrved velocity-integrated flux, $S^V$ (units of Jy km s$^{-1}$), we converted to an intrinsic brightness temperature luminosity, $L^T$ (units of K km s$^{-1}$ pc$^2$) following the derived equation in the appendix of \cite{obreschkow/etal:2009}:
\begin{align}
    \frac{L^T}{\rm K\; km\; s^{-1}\; pc^2} &= 
    3.255 \times 10^7
    \left( 
    \vphantom{\frac{A}{A}} 
        \frac{\nuobs}{\rm GHz} 
    \right)^{-2}
    \left( \frac{D_{\rm L}}{\rm Mpc} \right)^2
    (1+z)^{-3}
    \frac{S^V}{\rm Jy\; km\; s^{-1}}
\end{align}
From these galaxy luminosities, we generated CO intensity maps and power spectra as described in \S \ref{sec:genimapandpspec}.

\newpage

\begin{table*}
    \centering
    \caption{Previous CO Intensity Mapping Models}
    \begin{tabular}{l | c | p{0.33\textwidth}  p{0.33\textwidth} }
        \bf Paper
        &
        \bf Redshifts
        &
        \multicolumn{2}{c}{\textbf{CO luminosity model}}
        \tabularnewline
        \hline
        \hline

        \cite{righi/etal:2008}
        &
        $z \sim 0 - 10$
        &
        \textbf{Calculating SFR}: 
        SFR assumed to be driven by major mergers.  Stellar mass $M^*$ formed in a merger of total mass $M=M_1+M_2$ calculated as:
        \begin{align*}
            M^* = 4 \frac{\Omega_b}{\Omega_m} \eta \frac{M_1 M_2}{M} = 3.5 \times 10^{-2} \frac{M_1 M_2}{M}
        \end{align*}
        Merger rate was calculated from extended Press-Schechter formalism \citep{lacey/cole:1993}.
        \newline
        &
        \textbf{Converting SFR to $\bf \lco$}:
        Based on M82 observations \citep{weiss/etal:2005}:
        \begin{align*}
            \lco = 3.7 \times 10^3 \, SFR
        \end{align*}
        \\ \hline
        
        \cite{visbal/loeb:2010}
        &
        $z \gtrsim 3$
        &
        \textbf{Calculating SFR}:
        SFR proportional to halo mass, $M_{\rm halo}$:
        \begin{align*}
            SFR = 6.2\times 10^{-11} \left( \frac{1+z}{3.5} \right)^{3/2} M_{\rm halo}
        \end{align*}
        &
        \textbf{Converting SFR to $\bf \lco$}: 
        Based on M82 observations \citep{weiss/etal:2005}:
        \begin{align*}
            \lco = 3.7\times 10^3 \, SFR
        \end{align*}
        \\ \hline
        
        \cite{gong/etal:2011}
        &
        $z \sim 6 - 8$
        &
        \textbf{Calculating $\bf \lco$}:
        Fit a relation of the form
        \begin{align*}
            \lco
            = L_0 \left( \frac{M_{\rm halo}}{M_c} \right)^b
            \left( 1 + \frac{M_{\rm halo}}{M_c} \right)^{-d}
        \end{align*}
        to the results from semi-analytic modeling by \cite{obreschkow/etal:2009}.
        &
        At $z=6$, 7, and 8 respectively:
        {\begin{align*}
            L_0 &= 4.3 \times 10^6, \, 6.2 \times 10^6, \, 4.0 \times 10^6 \; \Lsun \\
            b   &= 2.4,\, 2.6,\, 2.8 \\
            M_c &= 3.5 \times 10^{11},\, 3.0 \times 10^{11},\, 2.0\times 10^{11} \; \Msun \\
            d   &= 2.8, 3.4, 3.3 
        \end{align*}}
        \\ \hline
        
        \cite{carilli:2011}
        &
        $z \sim 6 - 10$
        &
        \textbf{Calculating SFR}:
        Cosmic SFR density calculated as the value required to reionize the
        universe at a given redshift \citep{bunker/etal:2010, madau/etal:1999}.
        &
        \textbf{Converting SFR to $\bf \lco$}: 
        Combining SFR-$\lir$ \citep{kennicutt:1998} and disk galaxy $\lir$-$\lco$ \citep{daddi/etal:2010} relations:
        \begin{align*}
            \lco = 1.0\times 10^4 \, SFR
        \end{align*}
        \\ \hline
        
        \cite{lidz/etal:2011}
        &
        $z\gtrsim 6$
        &
        \textbf{Calculating SFR}:
        SFR proportional to halo mass, $M_{\rm halo}$:
        \begin{align*}
            SFR = 1.7\times 10^{-10}\, M_{\rm halo}
        \end{align*}
        &
        \textbf{Converting SFR to $\bf \lco$}: 
        Combining SFR-$\lir$ \citep{kennicutt:1998} and $\lir$-$\lco$ \citep{wang/etal:2010} relations:
        \begin{align*}
            \lco = 3.2 \times 10^4 \, SFR
        \end{align*}
        Note: as originally derived, $\lco \propto (SFR)^{3/5}$.  Linear scaling adopted for simplicity.
        
        \\ \hline
        
        \cite{pullen/etal:2013}
        &
        $z \sim 2 - 4$
        &
        \textbf{Calculating SFR}:
        Model $A$: SFR proportional to $(M_{\rm halo})^{5/3}$:
        \begin{align*}
            SFR = 1.2\times 10^{-11}\, (M_{\rm halo})^{5/3}
        \end{align*}
        
        Model $B$: SFR calculated from empirically fit SFR Schechter
        functions. \citep{smit/etal:2012}
        &
        \textbf{Converting SFR to $\bf \lco$}: 
        Same as \cite{lidz/etal:2011}:
        \begin{align*}
            \lco = 3.2 \times 10^4 \; (SFR)^{3/5}
        \end{align*}

        Note: for Model $A$, the effective model is still $\lco \propto M_{\rm
        halo}$.  For Model $B$, power spectra were obtained by
        multiplying those of $A$ by a factor $\langle T_{\rm CO,\, B}
        \rangle^2 / \langle T_{\rm CO,\, A} \rangle^2$ ($\approx 4.8^2$ at $z=3$).

    \end{tabular}
    \tablecomments{For consistent comparison, relations here were simplified
        using the authors' given fiducial assumptions, but we refer the reader
        to the original papers for full parameterizations and details.  See
        Figure \ref{fig:lco_v_mass} for a comparison of $\lco(M_{\rm halo})$
        between some of these models.
    }
    \label{tab:prevstudies}
\end{table*}

\end{appendix}

\end{document}